\documentclass[longbibliography,aps,prb,preprint,superscriptaddress]{revtex4-2}

\usepackage[latin9]{inputenc}

\usepackage{wrapfig}
\usepackage{units}
\usepackage{amsbsy}
\usepackage{amstext}
\usepackage{amsmath}
\usepackage{graphicx}
\usepackage{xcolor}
\usepackage{times}
\usepackage{float}
\usepackage{setspace}

\usepackage{hyperref}

\usepackage{amssymb}

\begin{document}

% ========================================Title

\title{Search for orbital magnetism in the kagome superconductor ${\rm CsV_3Sb_5}$ using neutron diffraction}

% Place the author information here.  Please hand-code the contact
% information and notecalls; do *not* use \footnote commands.  Let the
% author contact information appear immediately below the author names
% as shown.  We would also prefer that you don't change the type-size
% settings shown here.

% ========================================Authors

\author{William Li\`ege}
\affiliation{Universit{\'e} Paris-Saclay, CNRS, CEA, Laboratoire L{\'e}on  Brillouin, 91191, Gif-sur-Yvette, France}
\author{Yaofeng Xie}
\affiliation{Department of Physics and Astronomy, Rice University, Houston, Texas 77005, USA}
\author{Dalila Bounoua}
\affiliation{Universit{\'e} Paris-Saclay, CNRS, CEA, Laboratoire L{\'e}on  Brillouin, 91191, Gif-sur-Yvette, France}
\author{Yvan Sidis}
\affiliation{Universit{\'e} Paris-Saclay, CNRS, CEA, Laboratoire L{\'e}on  Brillouin, 91191, Gif-sur-Yvette, France}
\author{Fr\'ed\'eric Bourdarot}
\affiliation{Institut Laue-Langevin, 71 avenue des Martyrs, Grenoble 38042, France}
\author{Yongkai Li}
\affiliation{Centre for Quantum Physics, Key Laboratory of Advanced Optoelectronic Quantum Architecture and Measurement (MOE), School of Physics, Beijing Institute of Technology, Beijing, China.}
\affiliation{ Beijing Key Lab of Nanophotonics and Ultrafine Optoelectronic Systems, Beijing Institute of Technology, Beijing, China.}
\author{Zhiwei Wang}
\affiliation{Centre for Quantum Physics, Key Laboratory of Advanced Optoelectronic Quantum Architecture and Measurement (MOE), School of Physics, Beijing Institute of Technology, Beijing, China.}
\affiliation{ Beijing Key Lab of Nanophotonics and Ultrafine Optoelectronic Systems, Beijing Institute of Technology, Beijing, China.}
\author{Jia-Xin Yin}
\affiliation{ Department of Physics, Southern University of Science and Technology, Shenzhen, Guangdong, China.}
\author{Pengcheng Dai}
\affiliation{Department of Physics and Astronomy, Rice University, Houston, Texas 77005, USA}
\author{Philippe Bourges}
 \email{philippe.bourges@cea.fr}
\affiliation{Universit{\'e} Paris-Saclay, CNRS, CEA, Laboratoire L{\'e}on  Brillouin, 91191, Gif-sur-Yvette, France}

%%%%%%%%%%%%%%%%% END OF PREAMBLE %%%%%%%%%%%%%%%%

% Double-space the manuscript.

% \baselineskip24pt

% Make the title.
% Place your abstract within the special {sciabstract} environment.

% ========================================Abstract 
\begin{abstract}

\noindent  
As many Kagome metals, the topological superconductor AV$_3$Sb$_5$ with (A = K,Rb,Cs) hosts a charge density wave . A related chiral flux phase that breaks the time-reversal symmetry has been further theoretically predicted in these materials.  The flux phase is associated with loop currents that produce ordered orbital magnetic moments, which would occur at the momentum points, $\bf M$, characterizing the charge-density wave state.  Polarized neutron-diffraction experiments have been performed on an assembly of single crystals of  ${\rm CsV_3Sb_5}$ to search for such orbital magnetic moments.  No evidence for the existence of a three-dimensionally ordered moment is found at any temperature at the first {\bf M$_1$}=(1/2,0,0) point in the Brillouin zone within an excellent experimental uncertainty, {\it i.e.} ${\bf m}=0 \pm 0.01\mu_B$ per vanadium atom. However, a hint to a magnetic orbital moment is found in the second Brillouin zone at ${\bf M_2}$=(1/2,1/2,0) at the detection  limit of the experiment. Some loop currents patterns flowing {\it only} on vanadium triangles are able to account for this finding suggesting an ordered orbital magnetic moment of, at most, $\sim 0.02 \pm 0.01\mu_B$ per vanadium triangle. 
\end{abstract}

\maketitle

% ========================================Introduction
\clearpage

\section{\label{Intro} Introduction}

The search for orbital loop currents (LCs) is a long-standing quest in both topological and correlated quantum matter. This exotic state of matter has been put forward to explain the enigmatic pseudogap phase of high-temperature cuprate superconductors \cite{Chakravarty01,Varma20}. Over the last years, this idea has gained in momentum  primarily through the observation of an intra-unit-cell magnetism using polarized neutron diffraction (PND) \cite{Fauque06,Bourges21}.  The observed hidden order parameter and its associated thermodynamic signatures \cite{Shekhter13,Sato17} correspond to a magneto-electric state that breaks both parity and time-reversal symmetries as evidenced by different experiments \cite{Fauque06,Zhao17,Zhang18}. Similar measurements in the iridates indicate the existence of an orbital magnetism consistent with loop currents in SrIr$_2$O$_4$ \cite{Jeong17}. It is thought to exist in a wider class of quantum materials beyond superconducting cuprates and oxides. The clearest example in favor of orbital magnetism is the observation of anomalous Hall effect in twisted bilayer graphene \cite{Sharpe19},  in the absence of any spin magnetism. Other examples mix the orbital response with the spin magnetism as it is recently established in ${\rm Mn_3Si_2Te_6}$, where the control of the colossal magnetoresistence was demonstrated to arise from an exotic quantum state that is driven
by ab plane chiral orbital currents flowing along the edges of MnTe$_6$ octahedra\cite{Zhang22}.  In honeycomb lattices, such flux phases were further suggested to interact with Dirac ferimions and produce the quantum  anomalous hall effect  \cite{Haldane88}. 

The quasi-2D kagome lattice materials have gained the spotlight over the last decade because of their rich ground states stemming from their non-trivial band structure.  In particular, the quasi-2D kagome lattice compounds AV$_3$Sb$_5$ with (A = K,Rb,Cs)  represent a new family of kagome metals \cite{Ortiz19} with intertwined  topological charge density wave (CDW) order around T$_{CDW}\sim$ 93K in the Cs compound and superconductivity at low temperature (below $\sim$ 2.5 K) \cite{Ortiz20}. The topological CDW is associated with a Fermi surface instability at the {\bf M}-point of the hexagonal structure as observed by scanning tunneling microscopy (STM) \cite{Wang21,Jiang21}.  Using X-ray diffraction, the CDW in  CsV$_3$Sb$_5$ was later proven to be three-dimensional, with a 2 x 2 x 2 superstructure \cite{Li21} larger than the 2 x 2 x 1 superstructure inffered from STM.  There are however some recent claims that the three dimensional ordering of the CDW can be sensitive to stacking along the c-axis \cite{Ortiz21,Xiao23} and possibly to the way samples are prepared.  Further, resonant X-ray studies at high pressure demonstrate that the 2 x 2 x 1 CDW associated with the vanadium kagome sublattice coexists with a Sb $5p$-electron assisted 2 x 2 x 2 CDW\cite{Li22b}. Interestingly, neutron diffraction  on an assembly of CsV$_3$Sb$_5$ single crystals reported an associated lattice distortion in the CDW state where the magnitude of the 2 x 2 x 1 superstructure was found to be larger \cite{Xie22} in contrast to the X-ray diffraction result. Additionally, rotational symmetry breaking from $C_6$ to $C_2$ symmetry in the CDW state, suggesting a nematic phase, has been reported through  spectroscopic-imaging STM  \cite{Li22Rotation} and  birefringence measurements \cite{Xu22}. However, recent transport  elastoresistivity measurements indicate no breaking of the 2-fold rotational symmetry below the CDW state, meaning no sign of nematicity \cite{Liu24}.

These materials are further characterized by a large anomalous Hall conductivity\cite{Yu-Hall21}. Accordingly,  a chiral CDW has been first reported using STM in KV$_3$Sb$_5$ with modulated intensity of the different charge peaks \cite{Jiang21}. However, further scanning tunneling measurements that used spin-polarized tips show no trace of the chiral flux current phase in CsV$_3$Sb$_5$ \cite{Li22} within their experimental uncertainty. Despite these experiments, these materials represent new platforms to investigate the interplay between topology, unconventional superconductivity and strong electron correlations where localized spin magnetism from vanadium atoms is readily thought to be absent as concluded from the macroscopic magnetic susceptibilty \cite{Ortiz19}.  Consistently, a muon spin spectroscopy ($\mu$SR) experiment  in polycrystalline KV$_3$Sb$_5$  samples does not show any static long-range magnetism in KV$_3$Sb$_5$ \cite{Kenney21} and  neutron powder measurements concluded that short range antifferromagnetic ordering is absent \cite{Ortiz19}. Therefore, local moment spin magnetism from $d$-electrons does not show up in any magnetic measurements leaving open the possibility of weaker orbital magnetism.  

Different approaches were employed to study the mechanism of electronic instabilities and explain the charge ordering and superconductivity, leading to the emergence of nematic chiral charge order, charge bond order  and the appearance of orbital moments  \cite{Park21,Christensen21,Denner21,Feng21,Lin21,Zhou22,Christensen22}. Chiral flux phases, with orbital loop currents, have been theoretically predicted in these materials  \cite{Park21,Feng21,Lin21,Zhou22,Christensen22}. The proposed chiral flux state breaks time-reversal symmetry (TRS) 
and results in an anomalous Hall effect. That is consistent with the  observation of a huge enhancement of the anomalous Hall effect at the CDW temperature\cite{Yu-Hall21}, which in principle, requires TRS breaking in the CDW state. A few recent $\mu$SR studies report time-reversal symmetry breaking below the charge order temperature (around 80 -95K depending on the material)\cite{Mielke22,Yu21,Khasanov22,Guguchia23}. This is actually interpreted as originating from loop current states and is reminiscent of those reported in cuprates  \cite{Varma20,Bourges21}. For instance, Mielke et al \cite{Mielke22} propose a current pattern among the triangles of vanadium atoms but the exact LC pattern can only be deduced from diffraction techniques. Additionally, electronic magneto chiral anisotropy gives convincing evidence for the TRS breaking phase, although its onset temperature is still under debate \cite{Guo22}. Even more recently, an optical manipulation of the CDW state RbV$_3$Sb$_5$ suggests an unusual piezo-magnetic response that requires TRS breaking \cite{Xing24}.

 Consistently, torque magnetometry experiments report the existence of an odd-parity order parameter appearing at a temperature T$^*$=130K $>$T$^{CDW}$  above the CDW order temperature \cite{Asaba24}, suggesting that TRS is broken below  T$^*$.  Using symmetry arguments, hidden Dirac multipoles, involving anapoles, have been also predicted to occur in  AV$_3$Sb$_5$ \cite{Scagnoli22} to explain the muSR results. However,  recent resonant X-ray experiments \cite{Li22b,Scagnoli24} only show resonant enhancement at the Sb L-edge at the 2 x 2 x 2 superstructure and no resonant feature at the V K-edge. The resonant effect at the Sb L-edge is attributed to anisotropic contributions from the different Sb sites further away from the  kagome lattice. As a result, for both CDW superstructures, no particular sensitivity to magnetic contributions (which would lead to time-reversal breaking) was reported  \cite{Scagnoli24}, possibly due to the weakeness of the expected feature. Spectroscopic-imaging STM and angle-resolved photoemission spectroscopy reveal  as well the existence of small reconstructed Fermi pockets\cite{Li23}. They are due to Fermi surface reconstruction that is induced by the 2 x 2 x 1 CDW state that could acquire orbital magnetic moments if time-reversal symmetry is broken in the CDW state\cite{Zhou22}. 

The orbital magnetic moments can be detected using polarized neutron diffraction although the experiment is challenging due to the weakness of the expected orbital magnetism. Indeed, closed in-plane loops lead to magnetic moments pointing perpendicular to the hexagonal plane  A structure factor calculation suggests magnetic intensity at different points in Q-space depending on the correlations between loop currents in neighboring unit cells. For the chiral flux phase proposed in these materials\cite{Jiang21,Feng21,Zhou22}, the CDW modulates the interactions between unit cells, with a magnetic intensity expected at half-integer $H$ such as the {\bf M}-points where the CDW occurs \cite{Li21,Xie22}. 

In this paper,  we report a quantitative study of CsV$_3$Sb$_5$ single crystals with polarized neutron diffraction. Despite numerous efforts, we could not observe any reliable magnetic signal at the $\bf M$-point in the first brillouin zone, {$\bf M_1$}=(1/2,0,0), corresponding to the propagation wavevector of the unconventional chiral CDW in the frustrated kagome lattice. From the threshold of our statistical error bars, one can determine an upper limit of the magnetic moment which has to be lower than the experimental uncertainty of $\sim$ 0.01 $\mu_B$ per vanadium at this {\bf Q}-position.  Other magnetic patterns that preserve the lattice symmetry have been investigated  as well with no indication of magnetism. However, we found a hint of a weak magnetic signal at the M-point,  {$\bf M_2$}=(1/2,1/2,0), in the second brillouin zone. A conservative estimate of  the magnetic signal suggests a magnetic moment of, at most, $\sim$ 0.02$\pm 0.01 \mu_B$  per formula unit. Such a low value sounds consistent with a few Gauss internal field estimated at the muon stopping site from $\mu$SR experiments \cite{Mielke22}.  Our finding questions the different loop current models that have been discussed in the kagome superconductors. In particular, a certain number of loop currents patterns flowing {\it only} on vanadium triangles can account for our polarized neutron data, in contrast to LCs patterns flowing on hexagons.

\section{\label{Bg} Background}

\subsection{\label{cross-section}  Neutron cross section and polarization}

It is straightforward to consider that LCs phases can be tested by neutron diffraction as the closed circulating charge loops give rise to a magnetic field, in principle perpendicular to the loop.  In addition to the interaction between neutron and atomic nuclei, the neutron spin interacts with any source of magnetic fields present in the  materials. The only experimental limits are the available neutron flux and the sample mass. Usually, one considers the magnetic field ${\bf B}$ generated by unpaired electron spins ($S$). However, orbital currents contribute equally. The magnetic field probed  by the neutron spin at any point ${\bf R}$ can actually be decomposed in  two terms\cite{Squires,Lovesey} as: 
 
\begin{eqnarray}
{ \bf B (\bf{R})} &=   { \bf B_S} +  { \bf B_L} &= \frac{\mu_0}{4\pi} \large\{ 
2  \mu_B {\bf \nabla} \Big( \frac{ {\bf S} \times  {\bf \hat{R} }}{R^2} \Big) - 
I  \frac{  d{\bf l} \times  {\bf \hat{R} }}{R^2}
\large\}
\label{spin+orbit}
\end{eqnarray}

$\mu_0$ is the vacuum permeability, ${\bf \hat{R}}$ is a unit vector in the direction 
of ${\bf R}$ and  $I$ is current intensity along the current path $d {\bf l}$. 
The first term comes from the spin of unpaired electrons and 
the second one from their orbital motion. The scattering due to the loop 
currents belongs to the second category. 

The neutron intensity is given by the neutron cross section, which in case of 
elastic magnetic scattering and polarized neutron can then be written, in general, as \cite{Squires,Lovesey}:

\begin{eqnarray}
{\frac{d \sigma}{d \Omega}}\Big{\vert}_{mag} =& \sum_{\tau} \vert F_M({\bf Q}) \vert ^2 \delta({\bf Q} - {\bf \tau}) \nonumber\\
 \vert F_M({\bf Q}) \vert ^2 =& {r_0^2} f(\bf{Q})^2 \vert < \pm \vert {\bf \sigma}.{\bf B}({\bf Q}) \vert \pm,\mp > \vert ^2 
\label{eq0}
\end{eqnarray}

$F_M({\bf Q})$ is the magnetic structure factor at the momentum tranfer ${\bf Q}$.  ${\bf \tau}$ represents  the propagation wavevector of the LCs state  
where the magnetic signal is expected in momentum space.  ${\bf \sigma}$ are Pauli matrices describing the neutron spin which can take the 
two states: $|+>$ and $|->$. Spin-flip (SF) and non-spin-flip (NSF) cross-sections are defined when the neutron spin is flipped or conserved after interaction with the sample, respectively. The Pauli matrices are related to neutron moment  as ${\bf \mu_N} =-\gamma\mu_N {\bf \sigma}$  where $\mu_N$ is the nuclear magneton and $\gamma$=1.913 is the gyromagnetic ratio for the neutron spin. The prefactor $r_0 =0.54\ 10^{-12}$ cm includes all multiplying factors discussed so far and  corresponds 
to the neutron magnetic scattering length for a magnetic moment of 1 Bohr magneton  ($\mu_B$) ($r_0/\gamma$ is the classical radius of the electron). 
% and ${r_0^2}=290$ mbarn. 

$f(\bf{Q})$ in Eq. \ref{eq0} is the atomic magnetic form factor (by definition, $f(0)=1$). In the present case, it corresponds to the Fourier transform of magnetic atomic orbitals involved in the loop currents state, here associated with the vanadium orbitals on which the LCS are assumed to flow.  Because the kagome AV$_3$Sb$_5$ materials exhibits no spin magnetism, one considers the magnetic form factor of V$^{4+}$ atom which carries no spin. That form factor is tabulated in the crystallographic tables. We show in table \ref{TabfQ} the values of that form factor for relevant ${\bf Q}$-points of the hexagonal reciprocal lattice of CsV$_3$Sb$_5$. 

\begin{table*}[h]
\centering
\scriptsize
\def\arraystretch{1.5}
\begin{ruledtabular}
  \begin{tabular}{ccccc}
              Momentum position: $(H,K,L)$      &M$_1$=(1/2,0,0)        & M$_2$=(1/2,1/2,0)      & (1,0,0)       & (1,1,0)       \\ \hline
Q (\AA $^{-1}$)    &  0.66  &     1.143  &  1.32  & 2.287    \\ 
$f({\bf Q})$   &  0.97   &     0.92  &  0.89  & 0.72    \\ 
$f({\bf Q})^2$   &  0.95   &    0.85  &  0.80  & 0.52    \\ 
 \end{tabular}
  \end{ruledtabular}
\caption {Magnetic form factor of  V$^{4+}$ orbitals at different key positions in ${\bf Q}$-points of the CsV$_3$Sb$_5$ hexagonal reciprocal lattice }
\label{TabfQ}
\end{table*}

The key term in Eq. \ref{eq0} is ${\bf B}({\bf Q})$,  usually called the interaction vector \cite{Squires,Lovesey}, 
which is proportional to the Fourier transform  of the real space magnetic field distribution ${\bf B({\bf R})}$, at the  wave vector {\bf Q}. 
Following Eq. \ref{spin+orbit}, ${\bf B}({\bf Q})$ can be, in general,
written as a sum of the spin part with the spin moment,  ${\bf M_S}=2\mu_B \bf{S}$,  and the orbital part (for which we keep on purpose  the notation in terms of  currents),

\begin{equation}
{\bf B}({\bf Q}) = \sum_j \exp(-i  {\bf Q.r}_j) \big\{ {\bf \hat{Q}} \times {\bf M_S} \times  {\bf \hat{Q}} 
- i {I\over{2\mu_B}} \frac{{\bf \hat{Q}} \times d{\bf l}_j}{Q} \big\}
\label{Bq-all}
\end{equation}

where ${\bf \hat{Q}}={\bf Q}/Q$. The sum is made over all magnetic sites, $j$, within each {\it magnetic unit cell}. 
There are a few different ways to express the neutron cross section arising from the orbital part. It is generally assumed in neutron textbooks\cite{Squires,Lovesey}, that one can define an orbital moment,  ${\bf M_L}$, associated to the magnetic field distribution of the orbital part. In the low $\bf q$ limit,  the  orbital moment generally  corresponds to the atomic orbital moment\cite{Lovesey}.  This is misleading for the case of loop currents because the currents flow on more than one atomic orbital. Here, in particular, one would consider the vanadium hexagons or triangles of the kagome lattice. In cuprates \cite{Varma20,Bourges21}, the LCs involve hybridizations of copper $d-$ and oxygen $p-$orbitals. Clearly, the currents geometry plays an important role that impacts the momentum dependencies of the structure factor.  It is however useful to describe the LCs contribution by an orbital moment located at the geometric center of the closed loops. In such a case,  the magnetic cross section is taken to be of the same general form whatever the source of magnetism.  One then arrives to the simple form that the neutron interacts with a moment, ${\bf M= M_S+M_L}$, which is simply the sum of the spin moment and the orbital moment, 
and ${\bf B}({\bf Q})$ takes the simple form, 

\begin{equation}
{\bf B}({\bf Q}) = \sum_j \exp(-i  {\bf q.r}_j) ( {\bf \hat{Q}} \times  {\bf M_j} \times  {\bf \hat{Q}})
\label{Bq-spin}
\end{equation}

That expression fully defines the magnetic structure factor for a set of magnetic moments, 
${\bf M_j}$, in the unit cell.  One can define ${\bf m_{\perp}(Q)} = {\bf {\hat Q}}  \times  {\bf M_j} \times {\bf {\hat Q}}$ which stands for the Fourier transform of the magnetic moment distribution perpendicular to the unity vector ${\bf {\hat Q}}={\bf Q}/|\rm {\bf Q}|=(H,K,L)/|\rm {\bf Q}|$. ${\bf m_{\perp}(Q)}$  includes  the neutron orientation factor \cite{Lovesey,Squires}, that implies, as usual, that one  measures in neutron diffraction only the magnetic component perpendicular to the  wave vector ${\bf \hat{Q}}$,  owing to the dipolar nature of the magnetic scattering potential. 

\subsection{\label{calculation}  Calculation of the structure factor for different loop current patterns}

\begin{figure}
 \begin{centering}
 \includegraphics[width=9cm]{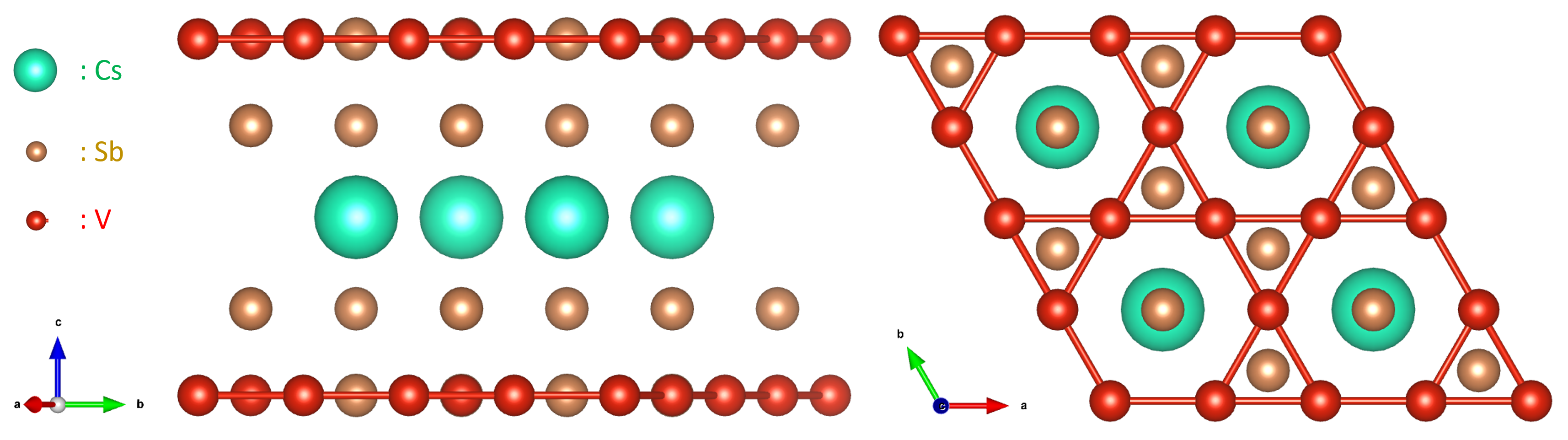}\\
 \includegraphics[width=5cm]{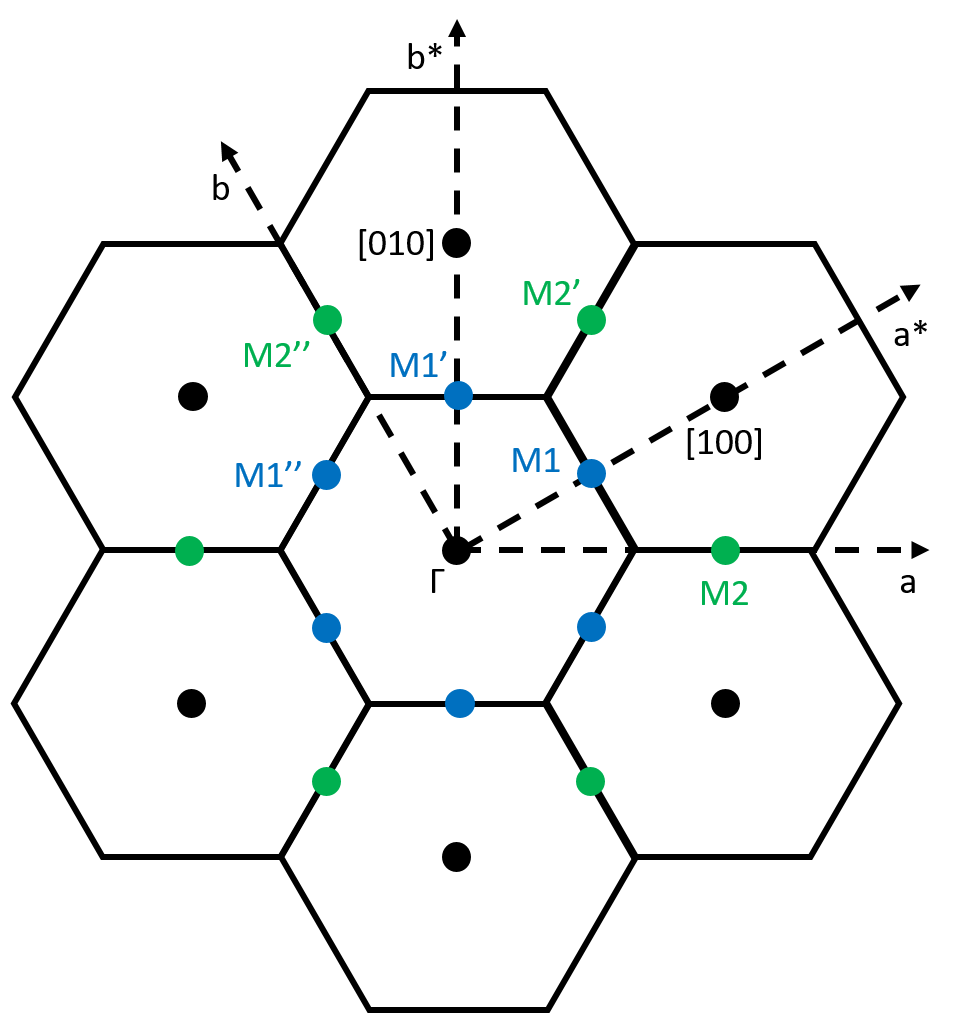} 
 \includegraphics[width=3cm]{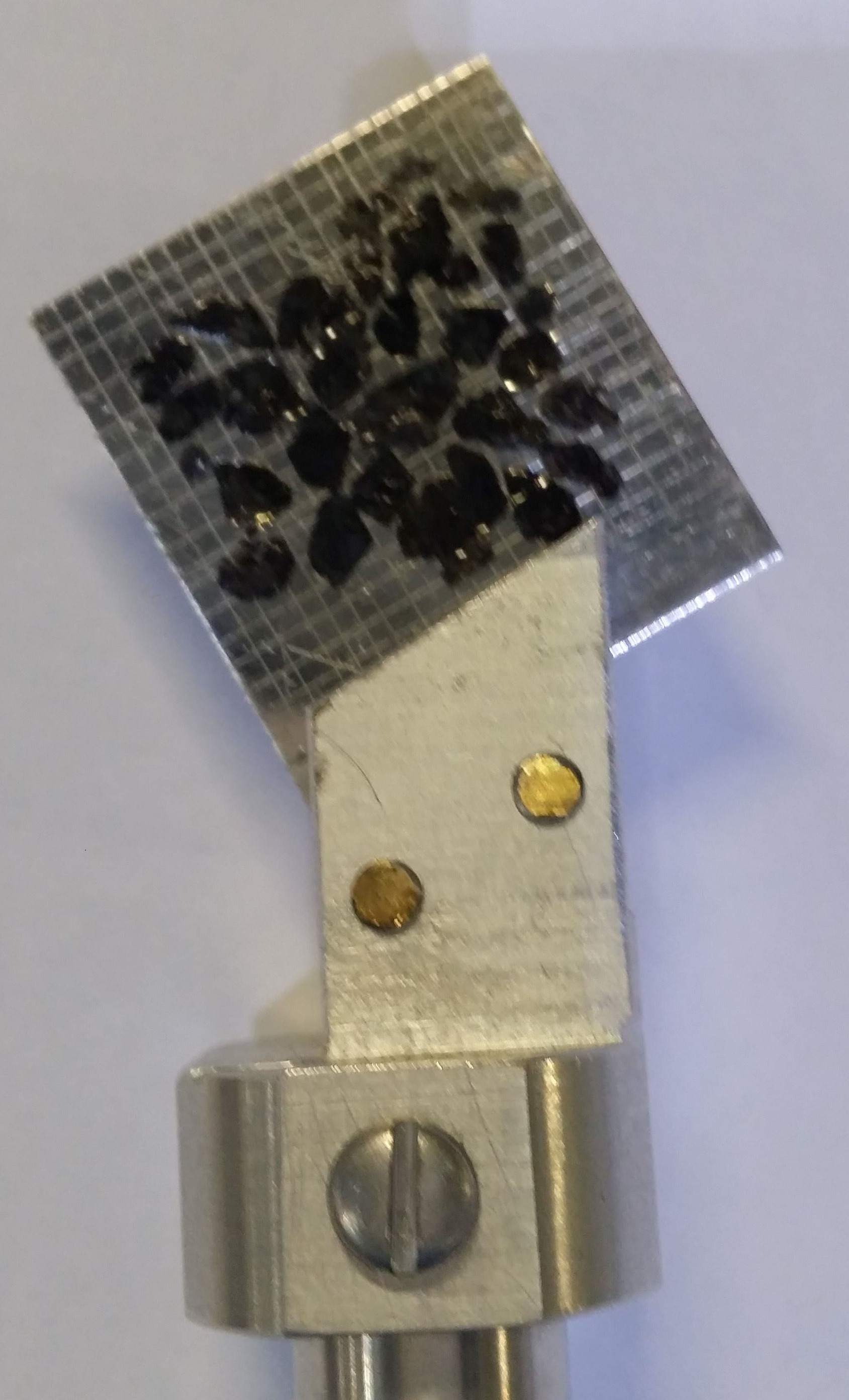} 
 \par\end{centering}
 \clearpage
\caption{\label{Fig1} \textbf{(a)} Stacking along the $c$-axis and projection in the $\bf (a,b)$ basal plane of the  hexagonal structure of ${\rm CsV_3Sb_5}$. \textbf{(b)} 2D Reciprocal space of the hexagonal latttice. The real space directions  $\bf (a,b)$ and reciprocal directions $\bf (a^*,b^*)$ are represented as well as the equivalent  ${\bf M_1}$ (in blue) and  ${\bf M_2}$ (in green) points.   \textbf{(c)} Photograph of the sample assembly mounted in the $(H,H,L)$ plane.   ${\rm CsV_3Sb_5}$ single crystals are glued on both faces of the Al plate. } %\textbf{(c)} polarization setup}
\end{figure}

A certain number of loop current states have been proposed to exist in the chiral charge density wave states \cite{Park21,Feng21,Lin21,Zhou22,Christensen22} of CsV$_3$Sb$_5$.  These states break time-reversal symmetry and  lattice translational symmetry as they all quadruple (2x2) the hexagonal unit cell shown in Fig. \ref{Fig1}.a. Examples of such LCs states are shown in Figs. \ref{Fig2}-\ref{Fig5}.  We here evaluate the magnetic structure factor, ${\bf B}({\bf Q})$,  for these different  LCs configurations.  Classically, the closed loops are forming an orbital moment pointing in principle perpendicularly to the hexagonal plane. This is the quantity that is experimentally probed in neutron diffraction experiments. 

Before going into the detailed description of the LCs patterns, let us stress that they are all supposed to be 2D with no correlation between the hexagonal planes. The structure factor is then assumed to be $L$-independent in these calculations. The correlation between the vanadium planes stacked along the c-axis would, of course, also play a role (See Fig. \ref{Fig1}.a). It may be either in-phase or out-of-phase leading to magnetic contributions at L integer (L=0) or half-integer (L=1/2) as argued from $\mu$SR data \cite{Yu21}. In this section, we consider only the scattered intensities in the 2D hexagonal plane. 

These 2D patterns can be classified in different types depending on how the different vanadium atoms are connected by charge currents within the hexagonal plane in the 2x2 unit cell. There are many possible orbital currents configurations in the kagome lattice, considering either the vanadium atoms form hexagon or triangle blocks. Loop currents can develop on the different links between vanadium atoms. The only constraint is that the current distribution within the 2x2 unit cell is such that there is a conservation of currents at each lattice node,
so that there is no net magnetic moment within the 2x2 unit cell. For each moment distribution, one can estimate the magnetic neutron structure factor in the 2D $(H,K)$ hexagonal plane.  The 2D reciprocal space of the atomic hexagonal cell is shown in Fig. \ref{Fig1}.b. The zone boundary {\bf M}-points  where the CDW occurs are represented. There are 6 equivalent points,   ${\bf M_1}=\pm(1/2,0,0) \equiv \pm(0,1/2,0) \equiv \pm (1/2,-1/2,0)$ in the first Brillouin zone and 6 other points ${\bf M_2}=\pm(1/2,1/2,0) \equiv \pm(1,-1/2,0) \equiv \pm (1/2,-1,0)$ in the next Brillouin zones. In the following, we consider various loop current models involving either the vanadium hexagons or mixing triangles and hexagons. 

\subsubsection{\label{hexagonsonly}  Loop current patterns involving vanadium hexagons}

The first example to be considered is when loop currents are running {\it only} on vanadium hexagons or  {\it only} on vanadium triangles although this has not been proposed in any theory, yet. We here explicitly consider the cases of currents running around the hexagons for a sake of clarity. Similar structure factors can be obtained when loop currents  concern  {\it only} the vanadium triangles that will be discussed in section \ref{Discussion}. There are 4 hexagons in the 2x2 unit cell (Fig. \ref{Fig2}.a): one can assume that two hexagons exhibit a current running clockwise and two others with currents running anti-clockwise: two hexagons then carry a moment $+{\bf m}_{hex}$ and the two others $-{\bf m}_{hex}$.  This model somehow reminds the antiferromagnetic spin ordering in cuprates. A simple calculation leads to a structure factor: 

\begin{equation}
 {\bf B(Q)}= 4 {m}_{hex}  \sin \pi H \sin \pi K 
\label{FM2x2}
\end{equation}

The magnetic intensity map  obtained from this structure factor multiplied by the vanadium form factor $f(Q)^2$ is shown in  Fig. \ref{Fig2}.b. For that phase, the $C_6$ axis symmetry of the hexagonal lattice is broken: in reciprocal space, meaning that only one of the three independent {\bf M}-points leads to magnetic scattering. The magnetic intensity is found to be maximum at M$_1$=(1/2,-1/2,0) and at M$_2$=(1/2,1/2,0) whereas zero intensity is found at equivalent points.  It is worth stressing right away that one cannot distinguish these three different orientations experimentally, since they are equivalent from the atomic structure point of view. Indeed, for the magnetic scattering, this would result in an averaged intensity map of the three equivalent points due to the existence of magnetic domains.

Inspired by the model of Lin and Nandkishore \cite{Lin21}, other LCs distribution on hexagons that respect the $C_6$ axis symmetry are also possible.  Quite similar to the theoretical approaches \cite{Lin21,Zhou22}, one hexagon can carry a moment  ${\bf m}_{hex}$ whereas the 3 others have only  $-{1\over3}{\bf m}_{hex}$  (Fig. \ref{Fig3}.a) . In the 2D $(H,K)$ hexagonal plane, this leads to a magnetic neutron structure factor that can be written as,

\begin{equation}
{\bf B(Q)}={4\over3} {m}_{hex} (1 - \exp^{i \pi (H+K)} \cos \pi H \cos\pi K)
\label{FM2x2b}
\end{equation}

  From this moment distribution, the neutron magnetic intensity map obtained from the squared structure factor multiplied by the vanadium form factor $f(Q)^2$ is represented in Fig. \ref{Fig3}.b. It exhibits again maxima at both ${\bf M_1}$=(1/2,0,0) and  ${\bf M_2}$=(1/2,1/2,0) which only differ by the form factor term. All the six equivalent ${\bf M_1}$ and ${\bf M_2}$ show the same intensity respecting the $C_6$ axis symmetry unlike the previous model. 

\subsubsection{\label{hexagonsandtriangles}  Loop current patterns involving vanadium hexagons and triangles}

The second type of patterns mix both  hexagons and triangles sub-blocks.  There were originally proposed by Feng {\it et al} \cite{Feng21}: they considered LCs with currents running around either given vanadium triangles or hexagons (Fig. \ref{Fig4}.a).  There is a complete cancellation between the hexagon magnetic moment ${\bf m}_{hex}$ and the magnetic moment on each triangle ${\bf m}_{tr}$: ${\bf m}_{hex}+ 2{\bf m}_{tr}=0$. From this moment distribution, one can also calculate the magnetic neutron structure factor in the 2D $(H,K)$ scattering plane. One simply obtains: 

\begin{equation}
 {\bf B(Q)}={m}_{hex} [1-\cos({4\pi\over3} (H-K))]
\label{FM-Feng21}
\end{equation}

The squared structure factor,  again multiplied by the vanadium form factor $f(Q)^2$ to get the neutron intensity map, gives the neutron intensity shown in  Fig. \ref{Fig4}.b where the strongest magnetic intensity is found at ${\bf M_1}$=(1/2,0,0) whereas no intensity is expected  at ${\bf M_2}$=(1/2,1/2,0). 
All the six equivalent ${\bf M_1}$ show the same intensity respecting the $C_6$ axis symmetry.

A third example is when loop currents occur around {\it all} sub-blocks either the hexagons or the triangles \cite{Lin21,Zhou22} (Fig. \ref{Fig5}). A flux exists even if  the currents are not closing in all sub-patterns.  This is allowed in quantum mechanics although that would give zero flux in a classical picture. In such a case,
the currents on the different arms of the triangles/hexagons are different, and the line integral of the current around any given triangle/hexagon does not vanish even when the arrows do not form a loop. For instance in ref. \cite{Lin21}, when the currents, indicated by coloured arrows, are not the same in the different links and do not seem to circulate, there is still a non-zero flux, although it gives rise to a weaker flux. In the 2x2 plaquette, the total flux over all sub-blocks is always zero.  Interestingly, in Lin and Nandkishore \cite{Lin21}, the  fluxes can be reassembled in four different groups, each having the same pattern as the model discussed above of Feng {\it et al} \cite{Feng21} (Fig. \ref{Fig4}.a). As a result, this leads to the same structure factor as the one shown in Fig. \ref{Fig4}.b.  In  ref. \cite{Zhou22}, the null sum of fluxes translates into $\Phi_1+3\Phi_2+2\Phi_3+6\Phi_4$=0 for all hexagons and triangles of the 2x2 plaquette (Fig. \ref{Fig5}.a). The resulting neutron intensity map, plotted in Fig. \ref{Fig5}.b, is similar to the previous example with stronger peaks at the ${\bf M_1}$ points and very weak intensity at the ${\bf M_2}$ positions. 

\section{\label{Exp} Experimental details}
\subsection{\label{sample} Sample preparation and characterization}

Single crystals of ${\rm CsV_{3}Sb_{5}}$ of less than half  a mm$^3$ size each used in the present study were grown by a self-flux method as described elsewhere\cite{Wang21}. Transport data for ${\rm CsV_{3}Sb_{5}}$, confirms
the existence of CDW order below T$_{CDW}$ = 95K and superconductivity below Tc $\simeq$ 2.5K \cite{Xie22}.
The single crystals were next glued and co-aligned onto aluminum plates with CYTOP (Amorphous Fluoropolymers) glue. About 400 individual single crystals were co-aligned on four aluminum plates to form an assembly with a volume of 0.11 cm$^3$ corresponding to a resulting total mass of $\sim$ 0.7 gram. The same sample assembly was previously used in our  phonon study with inelastic neutron scattering \cite{Xie22}.  

Two different mounting of the ${\rm CsV_{3}Sb_{5}}$ assembly were used to access the different {\bf M}-points of the Brillouin zone:  ${\bf M_1}$=(1/2,0,0) or ${\bf M_2}$=(1/2,1/2,0), allowing us to investigate different loop currents models. In both cases, we could access the $L$-dependencies to test possible doubling of the unit cell along $c^*$.   First, the whole sample array was oriented with $[H,0,0]$ and $[0,0,L]$ reciprocal lattice directions of the hexagonal crystal structure in the horizontal scattering plane to measure the ${\bf M_1}$=(1/2,0,0) point. In this scattering plane, the full crystal assembly was fitting inside a He cryostat. The sample mosaic spread was found to be 3.5deg as in our previous experiment \cite{Xie22}. In a second experiment, it was not possible to fit the whole assembly into the cryostat as we had to turn the aluminum plates by 30 deg to get the $[H,H,0]$ and $[0,0,L]$ reciprocal lattice directions in the horizontal scattering plane to be able measure the ${\bf M_2}$=(1/2,1/2,0) point. Only one aluminum plate with the more single crystals glued on it was used and turned by 30 deg (see sample photograph in Fig. \ref{Fig1}.c). The resulting mosaic spread was a bit better of $\sim$ 2.5 deg for a mass of about 40\% of the previous one.  As the sample is an assembly of smaller single crystals, the magnetic domains discussed above would be equally populated. In the following, the momentum transfer ${\bf Q}$ = H${\bf a^*}$ + K${\bf b^*}$ + L${\bf c^*}$ is denoted as $(H,K,L)$ in reciprocal lattice units (r.l.u.)  of the hexagonal lattice with a = b = 5.495 \AA, and c = 9.309 \AA.

\subsection{\label{Polar} Polarized neutron scattering experiments}

The polarized neutron measurements were performed on the neutron Triple Axis Spectrometers (TAS) IN22 located at Institut Laue Langevin (ILL) in Grenoble (France). The measurements were carried out in elastic conditions with the final neutron wavevector (k$_f$) equal to the incident one  (k$_i$).  The neutron wavevector was k$_f$=2.662 \AA$^{-1}$, yielding an energy resolution of about $\sim$1 meV. To get a clean beam, two Pyrolitic Graphite (PG) filters, installed on the scattered beam, were used to remove higher order harmonics. In particular, we put two PG filters to avoid second order harmonic from the nuclear peaks as we expect magnetic intensity at the zone boundary ($\bf M$-points) of the Brillouin zones. 

The beam was polarized and analyzed using Heusler crystals, Cu$_2$MnAl, with the (1,1,1) Bragg reflection. The instrument is equipped with longitudinal XYZ polarization analysis  (XYZ-PA) (see ref. \cite{Bounoua22}), a powerful technique to selectively probe and disentangle the magnetic response from the nuclear one with no assumptions on the background. Both SF and NSF scans were done in order to crosscheck the absence of nuclear scattering at magnetic positions. According to conventional notations, $\bf X$ stands for the direction of the neutron polarization parallel to the transferred momentum $\bf Q$, $\bf Y$ and $\bf Z$ polarization directions are both perpendicular to  $\bf Q$. $\bf Y$  is the in-plane orthogonal direction and $\bf Z$ is perpendicular to the scattering plane. The spin-flip, $SF_{X,Y,Z}$, and non-spin-flip, $NSF_{X,Y,Z}$, intensities for the different neutron polarizations were measured in order to be able to perform the XYZ polarization analysis. 

\begin{table}[h]
\centering{}%
  \begin{ruledtabular}
  \begin{tabular}{ccccccc} 
{  Setup   }& Sample & Bragg peak$\equiv(H,K,L)$ & {FR$_X$} & {FR$_Y$} &{FR$_Z$} \tabularnewline
\hline
{    CRYOPAD    }&{  CsV$_3$Sb$_5$   }&  (2,0,0) & ${16 \pm 0.15}$  & $16.1 \pm 0.2$ & $16.1 \pm 0.2$ \tabularnewline
\hline 
Helm  &{  CsV$_3$Sb$_5$   }& (1,1,0) & $15.9 \pm 0.1$    & $15.6 \pm 0.1$  & $ 16.1 \pm 0.1$   \tabularnewline
\hline 
Helm & Quartz & (1/2,1/2,0)  & $14  \pm 0.3$ & $13.8  \pm 0.4$ & $14.6  \pm 0.4$\tabularnewline
\end{tabular}
  \end{ruledtabular}
\caption{Summary of flipping ratios (FR)  measured on IN22 along the 3 directions  for both setups: Helm (meaning using the Helmholtz-like coils) or CRYOPAD. The flipping ratios were either measured on Bragg peaks of the sample or on the coherent scattering of a quartz specimen, where $(H,K,L)$ units are still given within the reciprocal lattice of  CsV$_3$Sb$_5$. Typical error bars on the flipping ratios are given for each setup. \label{TabFR}}
\end{table}

The experiments were performed in three different experimental runs using  two different setups to control the neutron polarization at the sample position. We utilized either an Helmholtz-like sets of coils or a  spherical polarization analysis device, CRYOPAD\cite{Cryopad}, with zero magnetic field at the sample chamber. In principle, CRYOPAD offers a better neutron polarization homogeneity when turning the polarization along the three different directions. However, both setups give very similar neutron polarization in all three directions.  The flipping ratios FR=$(\frac{NSF_{X,Y,Z}}{SF_{X,Y,Z}})$ for both setups were measured on the sample Bragg peaks as well as on a quartz specimen. The polarization is then found to be quite homogeneous along the three directions (FR=16) on Bragg peaks for both setups  (see the Table \ref{TabFR} for actual values). However, as usual, a slighly better polarization homogeneity is obtained with CRYOPAD compared to the Helmholtz-like coils setup. Lower flipping ratios are also found using the quartz specimen. This is expected to be the case as the amorphous quartz scatters at all scattering angles corresponding to a broader ${\bf Q}$-resolution compared to the Bragg peaks of CsV$_3$Sb$_5$, then leading to a more imperfect polarization. Even if the polarization is very homogeneous with tiny variations of the flipping ratio upon turning the polarization (see Tab. \ref{TabFR}), we apply corrections from imperfect neutron polarization as discussed in \cite{Stewart09}. This correction has negligible effects on the deduced magnetic intensity. 

\subsection{\label{polarization} Polarization analysis}

XYZ-polarization analysis (XYZ-PA) of polarized neutron diffraction allows us to distinguish between the nuclear and the magnetic contributions in the scattered intensity \cite{Squires,Lovesey}. For a nuclear scattering, the neutron spin remains unchanged and the scattered intensity is always measured in the NSF channel.  For the magnetic scattering, the intensity in each channel strongly depends on the direction of the neutron polarization $\bf{P_n}$ because neutron spins are described by Pauli matrices ${\bf \sigma}$ in Eq.  \ref{eq0}, whom quantization axis is set by $\bf{P_n}$. The scattered magnetic intensity is thus proportional to $\rm | {\bf \sigma}.{\bf m}_{\perp}(Q) | ^2$. As a result of the Pauli matrices, only the components of $ {\bf m_{\perp}(Q)}$ perpendicular to the neutron spin polarization vector $\rm {\bf P_n}$ are Spin-Flip,  $I_{mag}^{SF}({\bf P_n})\propto({\bf m}_{\perp})^{2}-({\bf m}_{\perp}.\,{\bf P_n})^{2}$, whereas the component $I_{mag}^{NSF}({\bf P_n})\propto({\bf m}_{\perp}.\,{\bf P_n})^{2}$ does not flip the neutron spin and remains in the NSF channel. To perform a standard XYZ polarization analysis, one then sets $\rm {\bf P_n}$ in  the laboratory referential with the three orthogonal unitary directions (X,Y,Z) defined above. 
%, where  $\rm {\bf P_n}$ is respectively parallel to $\rm {\bf {\hat Q}}$ (X), perpendicular to $\rm {\bf {\hat Q}}$ but still in the scattering plane (Y), and parallel to the vertical direction (Z). 

For the X polarization, ${\bf P_n}$ parallel to $\rm {\bf {\hat Q}}$, the full magnetic scattering is SF only.  Let us consider the case of a generic magnetic moment with  components along three cartesian directions. In the case of hexagonal lattice of the kagome metals, these directions correspond to either  $\bf (a,b^*,c)$, meaning {\it i.e.}  $\rm {\bf m}=({\bf m}_a,{\bf m}_{b^*},{\bf m}_c)$, or  $\bf (a^*,b,c)$. For LCs phases where there is no net magnetic moment, we are in the case where all magnetic moments are antiparallel to a single direction. In such a case,  the magnetic intensity measured in the SF channel, at a momentum position ${\bf Q}=(H,K,L)$, for the X polarization, $I_{mag}^{SF}(X)$, can be expressed  \cite{Lovesey}  as:

\begin{equation}
I^{SF}_{mag}(X) \propto  \left[1-|\frac{2\pi}{a}\frac{H}{{\bf Q}}|^2\right] {\bf m}_a^2 + \left[1-\frac{4}{3}|\frac{2\pi}{a}\frac{H/2+K}{{\bf Q}}|^2\right] {\bf m}_{b^*}^2 + \left[1-|\frac{2\pi}{c}\frac{L}{\bf Q}|^2\right] {\bf m}_c^2 
\label{ImagX}
\end{equation}

In contrast, the scattered intensities in the  two complementary Y and Z polarizations probe only sub-parts of this cross-section (Eq. \ref{ImagX}), depending on both i) the direction of the polarization and ii) the scattering plane of the given experiment, {\it i.e.} how the sample is oriented relative to the  (X,Y,Z)  referential. In the limit $L=0$, which is relevant for the discussed experiments,  $I^{SF}_{mag}(Z)$ probes the $c$-axis magnetic contribution, ${\bf m}_{c}^{2}$, which corresponds to the expected moment orientation for the LCs states discussed above. 

All these cross-sections are measured fot each polarization on top of a background as $I^{SF}_{meas}({\bf X,Y,Z})= I^{SF}_{mag}({\bf X,Y,Z})+I_{BG}({\bf X,Y,Z})$. $I_{BG}({\bf X,Y,Z})$ is assumed at first approximation to be independent of the polarization.  The total scattered magnetic intensity,  $I_{mag}$, is next extracted such as  (see \cite{Bourges18,Bounoua20,Bounoua22} for more details about the analysis of the neutron data):

\begin{equation}
I_{mag}= 2\, I^{SF}_{meas}({\bf X})- I^{SF}_{meas}({\bf Y}) -  I^{SF}_{meas}({\bf Z}) % \propto 
 \label{Imag}
\end{equation}

This expression is very useful as it gives access to the intrinsic magnetic intensity by removing the background contribution, simply by assuming the same background for the three polarizations. In practice, however, the background in the three polarizations can be sligthly different and the background contribution to $I_{mag}$ does not completely  cancel; it then may give rise to a shift from zero of the baseline even in the absence of magnetism. This effect can be particularly sizeable when the magnetic signal is very weak as we face in the present study. This is one of the most important challenges of such PND experiments. 

\subsection{\label{calibration} Intensity calibration}

To determine the absolute amplitude of the magnetic intensity, we need i) to estimate the magnetic structure factor in absolute units and ii) to take into account the instrument resolution. First, the magnetic scattering cross-section \cite{Squires,Lovesey},
$I_{mag}$, reads in absolute units: 

%--------------------
\begin{equation}
I_{mag}=\Phi_{S} \frac{N_{cell}}{V_0} R_0(Q_m) I_M({\bf Q})  = \Phi_{S}  \frac{N_{cell}}{V_0}   R_0(Q_m)  r_{0}^{2}\,f(Q)^{2}\, \frac{|B({\bf Q})|^{2}}{4}
\label{eq:Full equation}
\end{equation}

$\Phi_{S}$ corresponds to the neutron flux at the sample position in neutron/s/barns (1 barn= $10^{-24}$cm$^2$),
$N_{cell}$ is the number of unit cells (formula units) in the sample, $V_0$ is the volume of the unit cell, $I_M({\bf Q})=r_{0}^{2} f(Q)^{2} |{\bf  B({Q})}|^{2}/4$ where $r_{0}$ stands for the neutron magnetic scattering length, $r_{0}^{2}=290$ mbarn, $f(Q)$ is the magnetic form factor, ${\bf B(Q)}$ is the magnetic
structure factor defined in Eq. \ref{Bq-spin} and calculated in section \ref{calculation} for various LCs patterns in the 2x2 plaquette. As a result, one needs to divide  
$|B({\bf Q})|^2$  by 4 because $I_M$ is given per formula unit. The expected {\bf Q}-dependence of $I_M({\bf Q})$ is represented in  Figs. \ref{Fig2}-\ref{Fig5}. Assuming that the magnetic order of LCs is long-range, corrections from resolution effects are also included ar each ${\bf Q}$ position through the factor $R_0(Q_m)$ which normalised the triple-axis resolution function\cite{Dorner72} for a Bragg peak.  $R_0(Q_m)$ is readily calculated from instrument geometry and sample mosaicity. % \cite{Hennion}. 

% that contains the ${\bf m}_{\perp}^{2}$ term as discussed above by the right side of Eq. \ref{ImagX}. 

To estimate $I_M({\bf Q})$ in absolute units, one needs to calibrate the magnetic intensity by the nuclear intensity of a reference Bragg peak whose intensity $I_B$ can be written as a function of its nuclear structure factor, $F_N$, as

\begin{equation}
I_B =\Phi_{S}  \frac{N_{cell}}{V_0}  R_0(Q_N) {| F_N|^2}
\label{Bragg}
\end{equation}

$| F_N|^2$ can be obtained by a simple calculation of the structure factor or using structure factor calculation software like VESTA. It is expressed in barns. The magnetic intensity  in barns, $I_M({\bf Q})$, is simply deduced from Eqs. \ref{eq:Full equation} and  \ref{Bragg}, as

\begin{equation}
I_M({\bf Q})  = \frac{I_{mag}  }{I_B }  \frac{ R_0(Q_N) }{ R_0(Q_m) } {| F_N|^2} = r_{0}^{2} f(Q)^{2} \frac{|B({\bf Q})|^{2}}{4}
\label{calib}
\end{equation}

The left hand of Eq. \ref{calib} is determined experimentally whereas the right part is computed for the different LCs models. This calculated magnetic intensity is the quantity shown in all color map figures of the manuscript with a color scale given in mbarns assuming a magnetic moment of 0.02 $\mu_B$. 

\section{\label{results} Results}

We performed a series of polarized neutron diffraction on IN22 experiments to cover the various loop currents patterns discussed above in section \ref{calculation}. We had to mount the sample in different geometries to access the various points in ${\bf Q}$-space.

\subsection{\label{M1} Absence of magnetic scattering at the $\bf M_1$-point}

Following the proposed LCs models \cite{Park21,Feng21,Lin21,Zhou22,Christensen22}, we first  investigated the full ${\rm CsV_{3}Sb_{5}}$ sample in the  $(H,0,L)$ scattering plane as a magnetic intensity is systematically expected at $\bf M_1$ in the topological CDW state. The experiment has been performed in two separated runs. The run-1 was a quick measurement to cover different points in ${\bf Q}$-space whereas the second one focused on the possible magnetic scattering at the  ${\bf M_1}$=(1/2,0,0) and ${\bf Q}$=(1/2,0,1/2) points. The counting time for the second run was noticeably longer: typically 30 minutes/point for run-1 and up to 5 hours/point for run-2. The Figs. \Ref{Fig6}.a and b show for Run-1 the $L$-dependence at $H=0.5$ of both the non-spin-flip and spin-flip intensity for the three polarization states. A broad peak maximum at $L=0$ is observed in both channels. The spin flip intensity does not change with the polarization which indicates that this signal is {\it not magnetic} in origin within the error bars at any of these ${\bf Q}$ (at any $L$ value), in particular at (1/2,0,0), (1/2,0,1/2) and  (1/2,0,1) where a magnetic contribution could have been expected in the models. Similar measurements for $H$=1.5 (not shown) indicate the same trend.

 In Run-2, we repeated these measurements  to improve the statistics, namely the $L$-scan at $H=0.5$ but over a much limited L-range. The result in the spin-flip channel for a polarization along $X$ where the magnetic signal should be maximum is shown in Fig. \Ref{Fig6}.c. The scan performed at 5K did not show any structure. The scan exhibits a weak sloping background due to the sample geometry as the sample is made from an assembly of 4 different Al plates. 
Further, polarization analysis using Eq. \ref{Imag} is shown in Fig. \Ref{Fig6}.d  from which the magnetic intensity is readily extracted without assumptions on the background.  These two figures indicate no sign of a magnetic signal at  $\bf M_1$. 
However, it should be stressed that the baseline for the background  in Fig. \Ref{Fig6}.d is found  negative. As discussed above in section \ref{polarization}, this is related to small differences in the background contribution in the three polarizations that Eq. \ref{Imag} does not remove. In order to be able to give an upper limit of the magnetic intensity, we perform a statistical analysis of  Fig. \Ref{Fig6}.c and  Fig. \Ref{Fig6}.d. We fit these curves along $L$ by the following simple function, 

%--------------------
\begin{equation}
I=A \exp(-a L^2) + B + s L
\label{fitdata}
\end{equation}

The first term represents the possible magnetic intensity with an amplitude $A$ and a broadening factor $a$.  $a$ cannot be determined experimentally due to the limited number of points and is set to a value constrained by the sample mosaicity. The other terms are the background where the slope $s$ (not zero only for the $SF_X$ intensity of Fig. \Ref{Fig6}.c) is also constrained by the difference of the extreme points at $L=\pm0.5$. In the fit shown in  Fig. \Ref{Fig6}, there is therefore only two free parameters: $A$ and $B$. Fitting Fig. \Ref{Fig6}.c by Eq. \ref{fitdata} gives $A=0.2 \pm 3$.  Within error bars, no magnetic signal is observed in these two experiments at the $\bf M_1$-point as well as at the point $H=1/2$ and  $L$=1/4 or 1/2. This contrasts with the LCs theoretical predictions\cite{Feng21,Lin21,Zhou22} for which one expects the largest magnetic contribution at the  $\bf M_1$-point (see Figs. \ref{Fig3}-\ref{Fig5}).

\begin{table}[H]
\centering{}%
  \begin{ruledtabular}
  \begin{tabular}{ccccccc} 
{$(H,0,L)$}& $|F_N|^2$ (barns) &  $R_0 (Q_N)$  & $I_B$ & $R_0 (Q_M)$  & $I_{mag}$ (5K) & $I_M$ (mbarns) \tabularnewline
\hline
  (0,0,4) & 10.1 & 0.54 & 154000  & 1.78 & $<$3 & $<$ 0.06 \tabularnewline
\hline 
 (3,0,0)  & 11.3 & 0.35 & 71000 &1.78  &  $<$3  & $<$0.09 \tabularnewline
\end{tabular}
  \end{ruledtabular}
\caption{\label{absoluH0L}Calibration of the magnetic intensity at the ${\bf M_1}$=(1/2,0,0) point deduced from the nuclear Bragg peak $(H,0,L)$. The intensity in absolute units is given in barns per formula unit of ${\rm CsV_3Sb_5}$ as deduced from Eq. \ref{calib}.  The intensities are all expressed for a monitor M=1e6 corresponding to a counting time of 317 sec. The upper limit of the experimental magnetic intensity is estimated from the fit using Eq. \ref{fitdata} of  Fig. \ref{Fig6}.c.
 \label{TabcalibM1}}
\end{table}

Although no magnetism is observed, one can give an upper value of the magnetic moment  from these measurements. For that, we need to calibrate the neutron intensities by using nuclear Bragg peaks (see section \ref{calibration}). In the $(H,0,L)$ plane, one can use two different strong Bragg peaks for that purpose as listed in the table \ref{TabcalibM1}. The calibration of the intensity gives an upper limit for the magnetic intensity to be 0.075 mbarns on average that one conservatively rounds up to $\sim$ 0.1 mbarns and from which one can make in section \ref{Discussion} an estimate of the upper limit of the magnetic moment for different models. 

\subsection{\label{M2} Possible weak magnetic scattering at the $\bf M_2$-point}

Next, the sample was installed  on IN22 in a different  scattering plane, $(H,H,L)$, in order  to access $\bf M_2$=(1/2,1/2,0) and related  points. The  $L$-dependence of the spin-flip intensity is shown in Fig. \ref{Fig7}.a at 5K. As in the other scattering plane the data has been corrected for imperfect polarization. Although the counting time is similar to the data in the $(H,0,L)$ plane (Fig. \ref{Fig6}.c), the counts are weaker as only one Al plate (Fig. \ref{Fig1}.c), with ${\rm CsV_3Sb_5}$ samples glued on both faces, was mounted with a total sample mass of about $\sim$ 2/5 of the whole sample. We found a hint of a weak magnetic signal at the ${\bf M_2}$-point (1/2,1/2,0) in the  SF$_X$ $L$-scan at 5K (Fig. \ref{Fig7}.a). The reported magnetic signal is consistent with a long range three-dimentional ordering although one cannot prove that point due to the scarcity of data points. The polarization analysis at 5K still using Eq. \ref{Imag} for the same scan (shown in Fig. \ref{Fig7}.c)  suggests as well a magnetic contribution at the same point confirming the SF$_X$ scan.  Further, the same scan SF$_X$ at 110K (Fig. \ref{Fig7}.b) shows a reduction of the magnetic feature at  $\bf M_2$=(1/2,1/2,0). Again, this trend is confirmed by the polarization analysis at 110K where no feature at $L=0$ is observed (Fig. \ref{Fig7}.d). 
 A long counting time was necessary to achieve this result (about 5 hours/point for each polarization) underlying the limit and challenge of the experiment. At both temperatures, the reference line for the background is consistently shifted by a negative constant for all $L\ne0$ similar to what is found above for the $\bf M_1$ study (Fig. \ref{Fig6}.d). Relative to the background level, we are a bit more sensitive to such small effects here because the Helmholtz-like coils was used to control the neutron polarization where the background is less homogenous when turning  the polarization compared to CRYOPAD. Due to the weakness of the magnetic signal and the shift of the reference line, it is unfortunately not possible to extract with enough confidence the magnetic moment direction. 

\begin{table}[H]
\centering{}%
  \begin{ruledtabular}
  \begin{tabular}{ccccccc} 
{$(H,H,L)$}& $|F_N|^2$ (barns) &  $R_0 (Q_N)$  & $I_B$ & $R_0 (Q_M)$  & $I_{mag}$ (5K) & $I_M$ (mbarns) \tabularnewline
\hline
  (0,0,4) & 10.1 & 0.54 & 77785  & 1.09 & 3.6 $\pm$ 1.5  &  0.23 $\pm$ 0.11  \tabularnewline
\hline 
 (1,1,0)  & 11.3 & 0.62 & 89065 & 1.09 &  3.6 $\pm$ 1.5   &  0.26 $\pm$ 0.11 \tabularnewline
\end{tabular}
  \end{ruledtabular}
\caption{Calibration of the magnetic intensity at the ${\bf M_2}$=(1/2,1/2,0) point deduced from the nuclear Bragg peak $(H,H,L)$. The intensity in absolute units, given in barns per formula unit of ${\rm CsV_3Sb_5}$, is deduced from Eq. \ref{calib}.  The intensities are expressed for a moniteur M=1e6 corresponding to a couting time of 235 sec. The magnetic intensity estimation is obtained from the fits using Eq. \ref{fitdata} of  Figs. \ref{Fig7}.a and c. \label{TabcalibM2}}
\end{table}

Following the procedure discussed above for ${\bf M_1}$, we fit the data with Eq. \ref{fitdata} with only two free parameters. Fitting Fig. \Ref{Fig7} gives on average $A=3.6 \pm 1.5$ at 5K and  $A=1 \pm 1.5$ at 110K. Again, one can calibrate the neutron intensities to estimate the absolute amplitude of the possible magnetic intensity  using strong nuclear Bragg peaks and in the hypothesis of a long range three-dimentional magnetic ordering. In the $(H,H,L)$ plane, one can use two different strong Bragg peaks for that purpose as listed in the table \ref{TabcalibM2}. A conservative value of the magnetic scattering is then about 0.25 $\pm$ 0.11 mbarns from which one can make in section \ref{Discussion} an estimate of a  magnetic moment for the different models. 

\subsection{\label{Tdep} Temperature dependence of the spin flip intensity $\bf M_1$-point and $\bf M_2$-point.}

To better characterize the weak magnetic ordering, we study the temperature dependencies of the spin flip intensity at both $\bf M_1$ and $\bf M_2$-points. In particular, according to torque magnetometry experiments, TRS is broken at a temperature T$^*$=130K $>$T$^{CDW}$  with a maximum effect above $\sim$ 80K \cite{Asaba24} leaving open the possibility of a magnetic intensity maximum  around 100K at the $\bf M_1$-position which would allow us to reconcile with the theoretical predictions of a stronger  magnetic signal at ${\bf M_1}$. Fig. \ref{Fig8}.a represents  the SF and NSF intensity for a polarization along $X$ at the ${\bf M_1} = (1/2,0,0)$ as a function of temperature.  One notices a decreasing of the raw intensity upon warning similar in both channels which are simply scaled by a factor 2 at all temperatures.  Additional measurements for the polarizations along $Y$ and $Z$ (not shown) indicate that the signal is not magnetic at any temperature.  One then does not observe any maximum in the SF intensity in the range 80-130K as it would be expected from the reported torque magnetometry data \cite{Asaba24}.

Fig. \ref{Fig8}.b  shows the temperature evolution in the spin-flip channel at ${\bf M_2} = (1/2,1/2,0)$ for a polarization along $X$ where the magnetism is maximum. One may notice a decreasing of the intensity above $T_{CDW}$. For a comparison, we plotted the nuclear intensity due to the lattice distortion related with the CDW that was observed at large Q  on the same sample with unpolarized neutron at the position ${\bf Q}$=(3.5,0,0) below $T_{CDW}$ \cite{Xie22}.   Unfortunately, our current temperature dependence is not accurate enough to make any statements concerning any anomaly at T'$\sim$ 30K as it has been reported in the $\mu$SR experiments\cite{Mielke22,Yu21,Khasanov22,Guguchia23} or a continuation of the intensity up to T$^*$as seen in torque magnetometry measurements \cite{Asaba24}. 

\subsection{\label{Gamma} No magnetic scattering at the $\Gamma$-point}

 All the examples discussed so far are LC states corresponding to {\bf q} = 1/2 antiferromagnetic (AFM) order, i.e., magnetic order where the ordering propagation vector is the same as the CDW state. So, for completeness, we checked as well the possibility of a magnetic scattering at  {\bf q} = 0 which respects the lattice symmetry, like the LCs phase predicted and observed in cuprates \cite{Varma20,Fauque06,Bourges21}. We then consider patterns which respect the hexagonal lattice symmetry. It should be stressed that this has not been proposed theoretically in kagome metals. Anyway, it corresponds to the case of orbital currents decorating all triangles of the  kagome lattice with currents running on two vanadium triangles of a single unit cell in clockwise or anticlockwise directions (Fig. \ref{Fig9}.a) quite similarly to the case of cuprates \cite{Bourges21}. In such a case, the structure factor reads  $ { \bf B( Q)}=2 {m}_{tr} \sin({\pi\over3} (H-K))$. In this case, the magnetic intensity, shown in Fig. \ref{Fig9}.b, is zero at all {\bf M}-points and is maximum at  Bragg positions with integer $(H,K,L)$, like ${\bf Q}=(1,0,0)$.

Fig. \ref{Fig9}.c  shows the temperature dependence of both SF and NSF intensities at the very weak nuclear Bragg peak  ${\bf Q}$=(1,0,0) (about 10$^{-4}$ weaker than the strong Bragg peaks listed in Tab. \ref{absoluH0L}). No sizeable additional intensities  are observed in the SF channel  at low temperature compared to the reference intensity at 100K  (above the CDW ordering temperature which is about 94K in that sample\cite{Xie22}). This indicates no magnetic contribution in agreement with a polarization analysis (Fig. \ref{Fig9}.c) where also no difference of intensities is observed between the three different polarizations, XYZ. Other momentum positions like (1,0,1/2), (1,0,1) or even ${\bf Q}$=(1,1,1) in the $(H,H,L)$ scattering plane were also measured using XYZ polarization analysis. No sign of magnetism was evidenced at any of these momenta within our experimental accuracy. As the Bragg peaks at {\bf Q}=(1,0,0) and {\bf Q}=(1,1,1) are very weak, the accuracy of these measurements is quite good due to the absence of any polarization leakage. For the Bragg position ${\bf Q}$=(1,0,0), $R_0=0.97$ and with the estimate  $I_{mag}<5$cnts for a monitor M=1e6 (Fig. \ref{Fig9}.c), one gets an average of $I_M< 0.25$ (mbarns) to be  compared with about 1-2 mbarns observed in cuprates \cite{Bourges18}. This leads to a moment of less than 0.03 $\mu_B$  per vanadium triangle corresponding to the model discussed above and using the form factor given in Tab. \ref{TabfQ}. 

\section{\label{Discussion} Discussion}

The main finding of our polarized neutron diffration experiments in  ${\rm CsV_3Sb_5}$  is that the magnetic signal, if it exists, is unfortunately very weak, being actually at the limit of the current best experimental  PND capabilities. Although the reported magnetic intensity is very low, it is worth emphasizing that we do not meet here the experimental complications, faced for the study of {\bf q}=0 intra-unit-cell magnetism in cuprates\cite{Bourges18}, related to the neutron polarization inhomogeneities, as the magnetic signal is expected in kagome metals at the Brillouin zone boundaries where no nuclear Bragg peaks are present. The experimental challenges are instead rather the available neutron flux and sample mass. Therefore, the fact that we observe a noticeable difference between both scattering planes  enables one to draw meaningfull conclusions. 

\begin{figure}
 \begin{centering}
 \includegraphics[width=14cm]{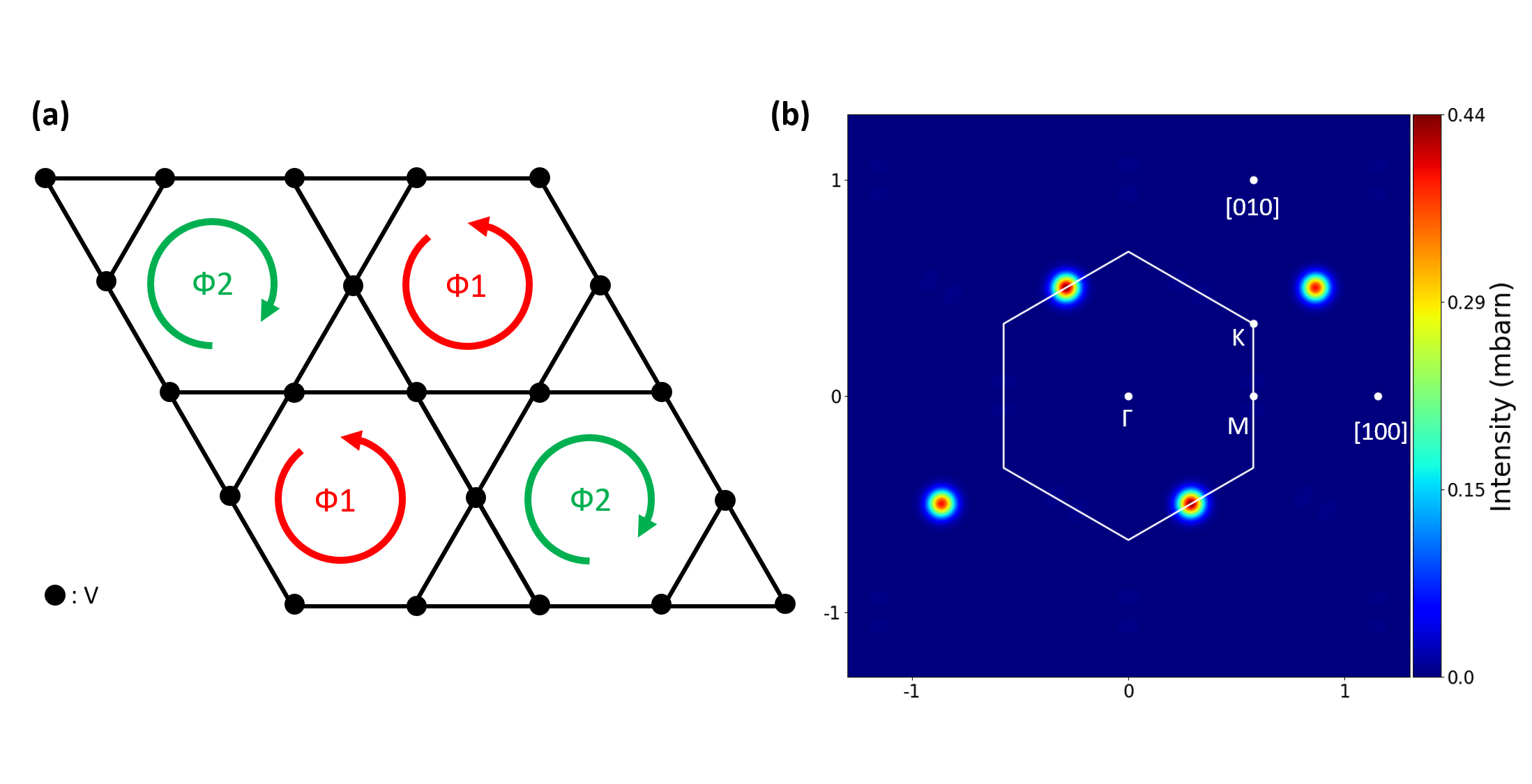} 
 \par\end{centering}
 \clearpage
\caption{\label{Fig2}  \textbf{(a)}  LCs with only moment on hexagons in the 2x2 unit cell for an antiferromagnetic model that breaks $C_6$ rotational symmetry with $\Phi_1+\Phi_2$=0. \textbf{(b)}   Calculated magnetic intensity in reciprocal space with similar intensities at both ${\bf M_1}$ and  ${\bf M_2}$ point. Due to the broken $C_6$ rotational symmetry,  magnetic intensities are not expected to be the same at all equivalent ${\bf M}$-points. As for all figures where the magnetic intensity is calculated, it is given in absolute units (mbarn) for a single unit cell using Eq. \ref{calib} assuming an ordered moment associated with  $\Phi_1$ of 0.02 $\mu_B$. }
\end{figure}

\begin{figure}
 \begin{centering}
 \includegraphics[width=14cm]{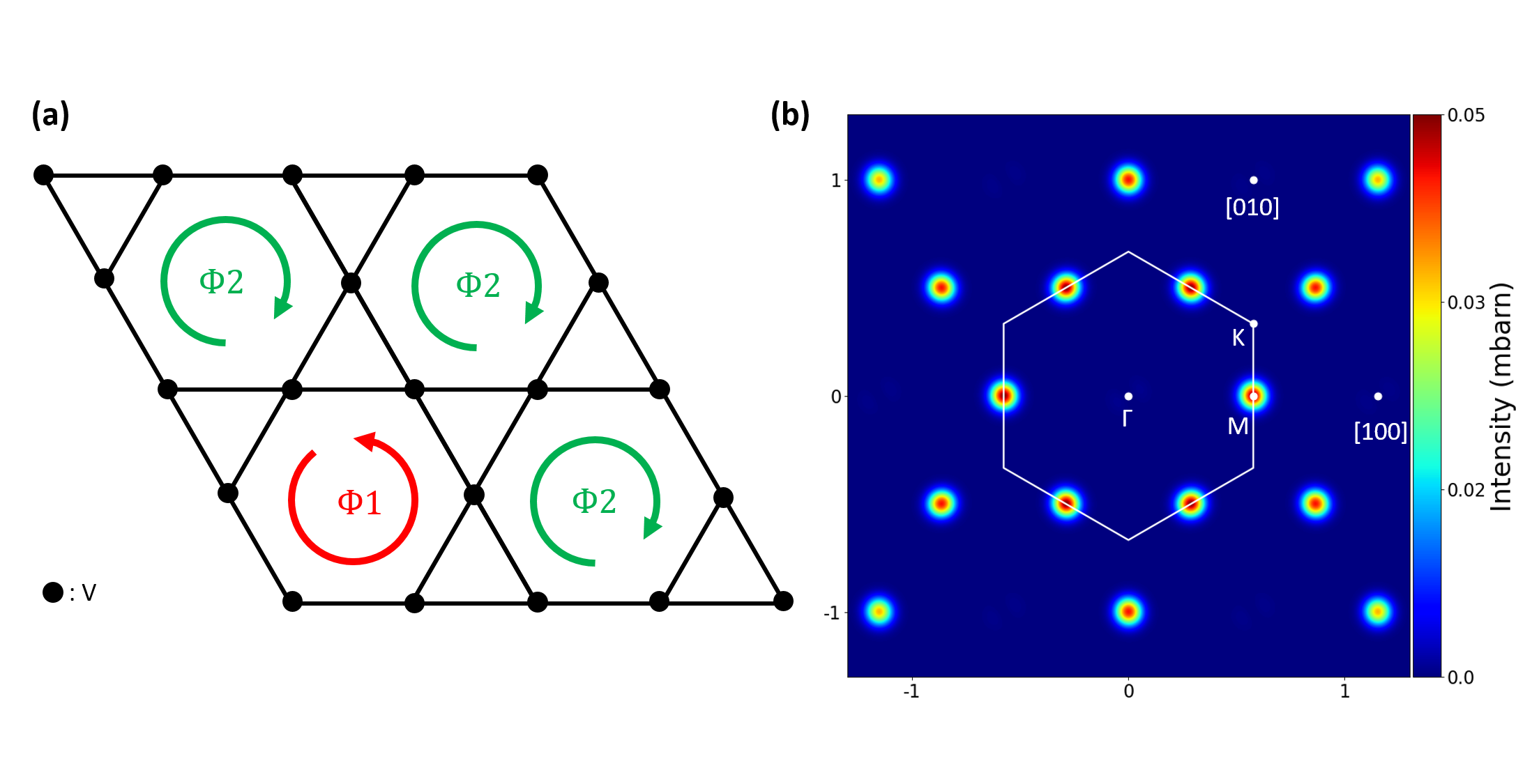}
 \par\end{centering}
 \clearpage
\caption{\label{Fig3}   \textbf{(a)} LCs configuration with moments only on hexagons in the 2x2 unit cell with a symmetry similar as the  models  proposed \cite{Lin21,Zhou22} for the LCs in kagome metals with  $\Phi_1+3\Phi_2$=0. \textbf{(b)}   Calculated magnetic intensity in reciprocal space with similar intensities at both ${\bf M_1}$ and  ${\bf M_2}$ points. To  calculate the magnetic intensity in absolute units, the magnetic moment associated with $\Phi_1$ is set to 0.02 $\mu_B$. }
\end{figure}

\begin{figure}
 \begin{centering}
 \includegraphics[width=14cm]{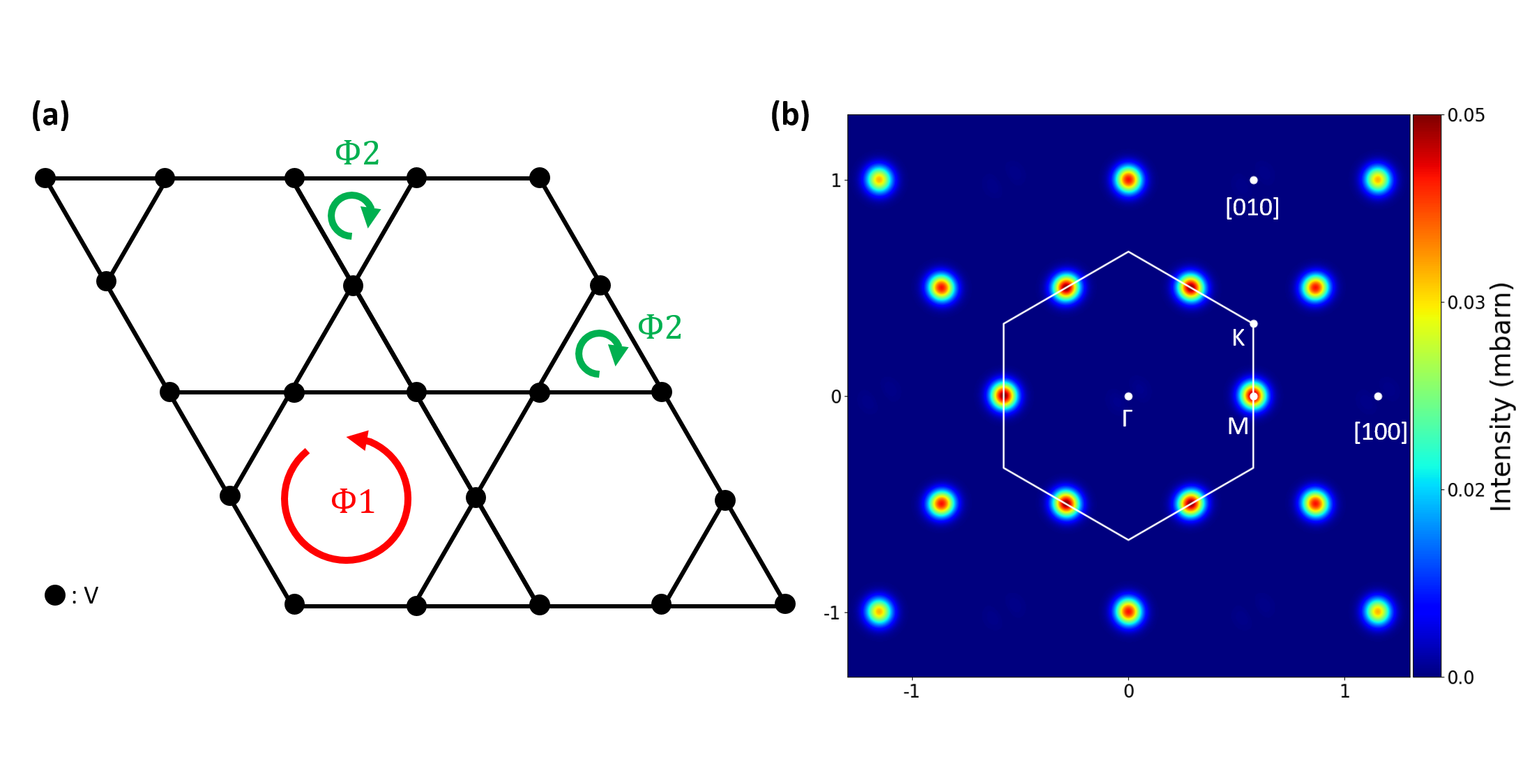} 
 \par\end{centering}
 \clearpage
\caption{\label{Fig4}    \textbf{(a)} LCs pattern in the 2x2 unit cell proposed in the model of Feng {\it et al} \cite{Feng21} with  $\Phi_1+2\Phi_2$=0.  \textbf{(b)}   Calculated magnetic intensity in reciprocal space with the maximum  intensity at  ${\bf M_1}$ and  zero intensity  ${\bf M_2}$. Additional intensities are observed at the Bragg position such as (1,0,0) and equivalent points.  To calculate the magnetic intensity in absolute units, the magnetic moment associated with $\Phi_1$ is set to 0.02 $\mu_B$.}
\end{figure}

\begin{figure}
 \begin{centering}
 \includegraphics[width=14cm]{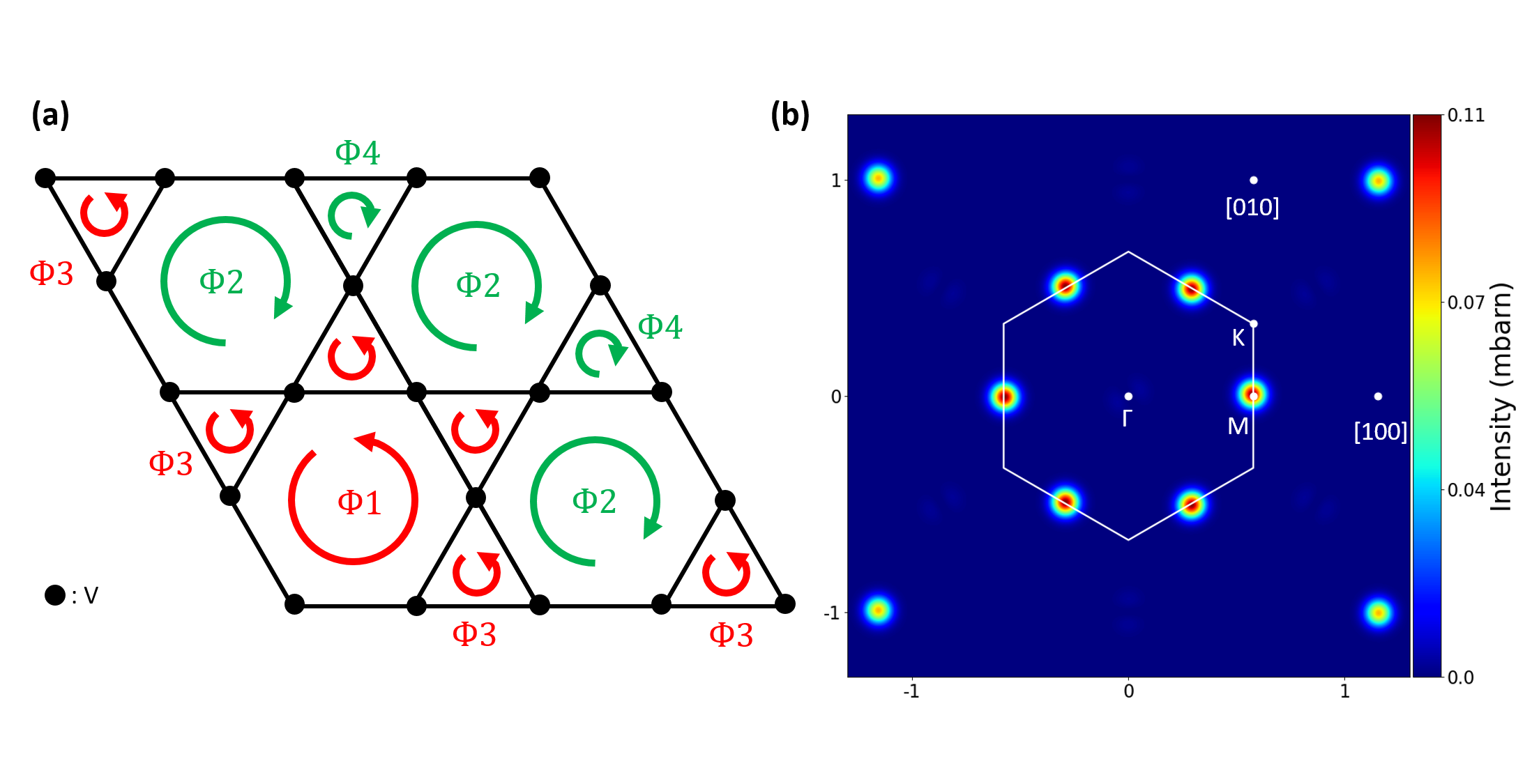}
 \par\end{centering}
 \clearpage
\caption{\label{Fig5}   \textbf{(a)} LCs configuration with moments only on hexagons in the 2x2 unit cell with a symmetry corresponding to the  proposed models \cite{Lin21,Zhou22} for the LCs in kagome metals.  \textbf{(b)}  Calculated magnetic intensity in reciprocal space with the maximum  intensity at  ${\bf M_1}$ and nearly  zero intensity  ${\bf M_2}$. To calculate the magnetic intensity in absolute units, the largest magnetic moment associated with $\Phi_1$ is set to 0.02 $\mu_B$.}
\end{figure}

\begin{figure}
 \begin{centering}
 \includegraphics[width=7cm]{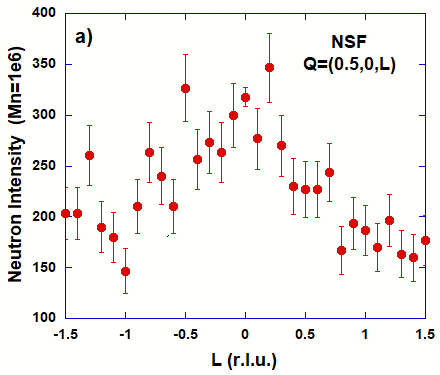} 
 \includegraphics[width=7cm]{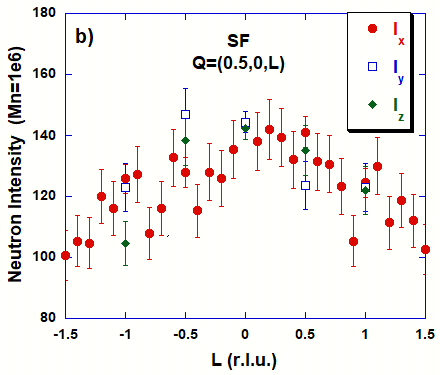} 
 \includegraphics[width=7cm]{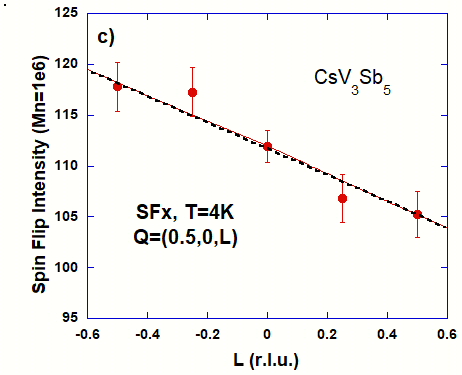} 
 \includegraphics[width=7cm]{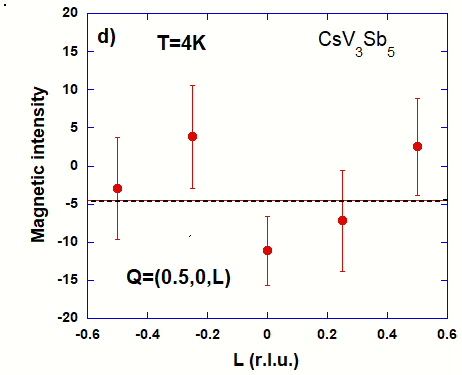} 
 \par\end{centering}
 \clearpage
\caption{\label{Fig6}  \textbf{L-dependence of the neutron intensity around ${\bf M_1}$ at T=4 K}:   \textbf{(a)} (Run-1) Scan along ${\bf Q}=(0.5,0,L)$  in the non-spin-flip channel $NSF_{X}$  \textbf{(b)} (Run-1) Scans  along ${\bf Q}=(0.5,0,L)$ in the spin-flip channel for the 3 polarizations.
 \textbf{(c)} (Run-2) Scans  along ${\bf Q}=(0.5,0,L)$ in the spin-flip channel for the $X$ polarization.   \textbf{(d)} (Run-2)  Magnetic intensity obtained by polarization analysis (Eq. \ref{Imag}).  The data in \textbf{(c-d)} have been corrected from imperfect polarization. The lines in  \textbf{(c-d)} are fits using Eq. \ref{fitdata} as explained in the text. The intensities are all reported for a monitor M=1e6 corresponding to a counting time of 317 sec although the counting time of the Run-1 experiment was about 30 minutes/points whereas it was about 5 hours/point in the Run-2 experiment.   In \textbf{(d)}, the reference line  for zero magnetic signal is negative (-4.8 cnts) due to inhomogeneous background for the different polarizations as explained in the main text.}
\end{figure}

\begin{figure}
 \begin{centering}
 \includegraphics[width=7cm]{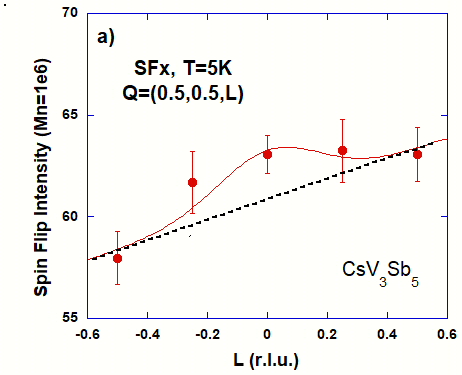} 
 \includegraphics[width=7cm]{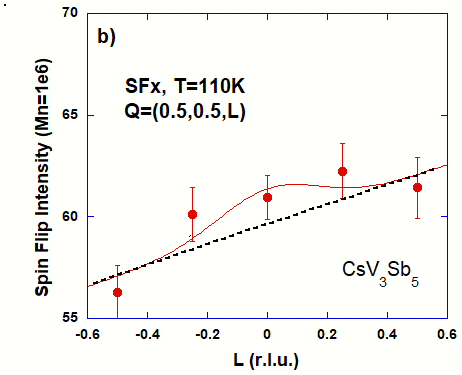} \\
 \includegraphics[width=7cm]{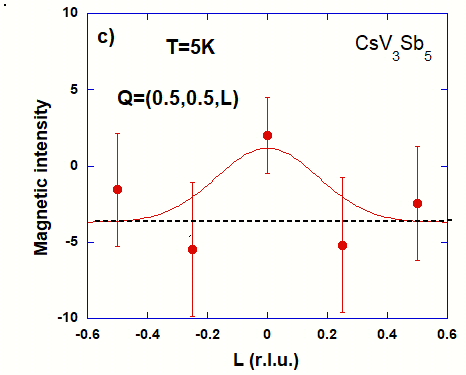} 
 \includegraphics[width=7cm]{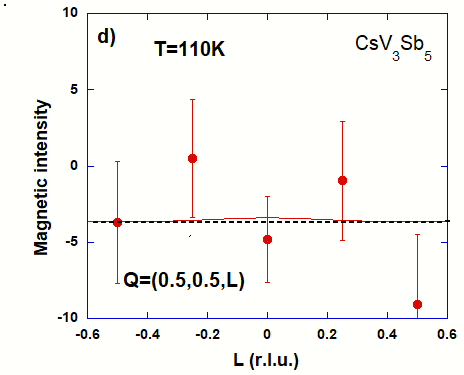} 
 \par\end{centering}
 \clearpage
\caption{\label{Fig7}  \textbf{L-dependence of the magnetic intensity at {\bf Q}=(1/2,1/2,L)}:   \textbf{(a-b)}  Scans along ${\bf Q}=(0.5,0.5,L)$  in the spin-flip channel $SF_{X}$ in ${\rm CsV_3Sb_5}$ at  \textbf{(a)} T=5K  and \textbf{(b)} T=110K.  \textbf{(c-d)}  Magnetic intensity obtained by polarization analysis (Eq. \ref{Imag}) at  \textbf{(c)} T=5K and  \textbf{(d)} T=110K. The data in \textbf{(a-d)} have been corrected from imperfect polarization. The lines in  \textbf{(a-d)} are fits using Eq. \ref{fitdata} as explained in the text. The intensities are all reported for a monitor M=1e6 corresponding to a counting time of 235 sec although the total counting time reaches about 5 hours/point for each polarization. In \textbf{(c-d)},  the reference line  for zero magnetic signal is negative due to inhomogeneous background for the different polarizations as explained in the main text.
That corresponds to a consistent shift by -3.6 cnts at both temperatures for $L\ne0$  for zero magnetic signal.   }
\end{figure}

\begin{figure}
 \begin{centering}
 \includegraphics[width=7.8cm]{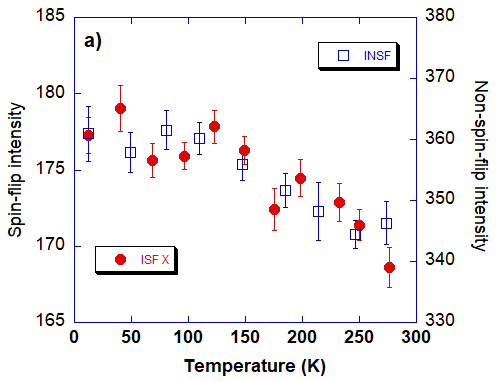} 
 \includegraphics[width=7cm]{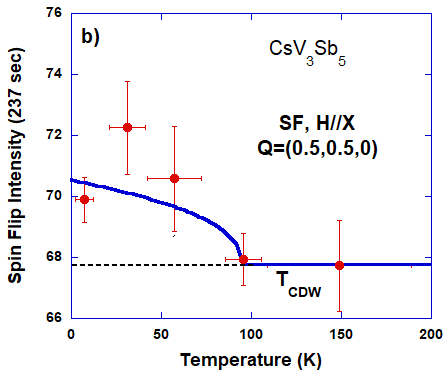} 
 \par\end{centering}
 \clearpage
\caption{\label{Fig8} \textbf{(a)} Temperature dependence of the SF (left scale) and NSF (right scale) intensities  at ${\bf M_1}$=(0.5,0,0) for a polarization along $X$.  
 \textbf{(b)} Temperature dependence of the spin flip intensity  at ${\bf M_2}$=(0.5,0.5,0) for a polarization along $X$. The fit of the  temperature evolution of the CDW signal observed in neutron diffraction on the same sample\cite{Xie22} is scaled for a comparison.}
\end{figure}

\begin{figure}
 \begin{centering}
\includegraphics[width=14cm]{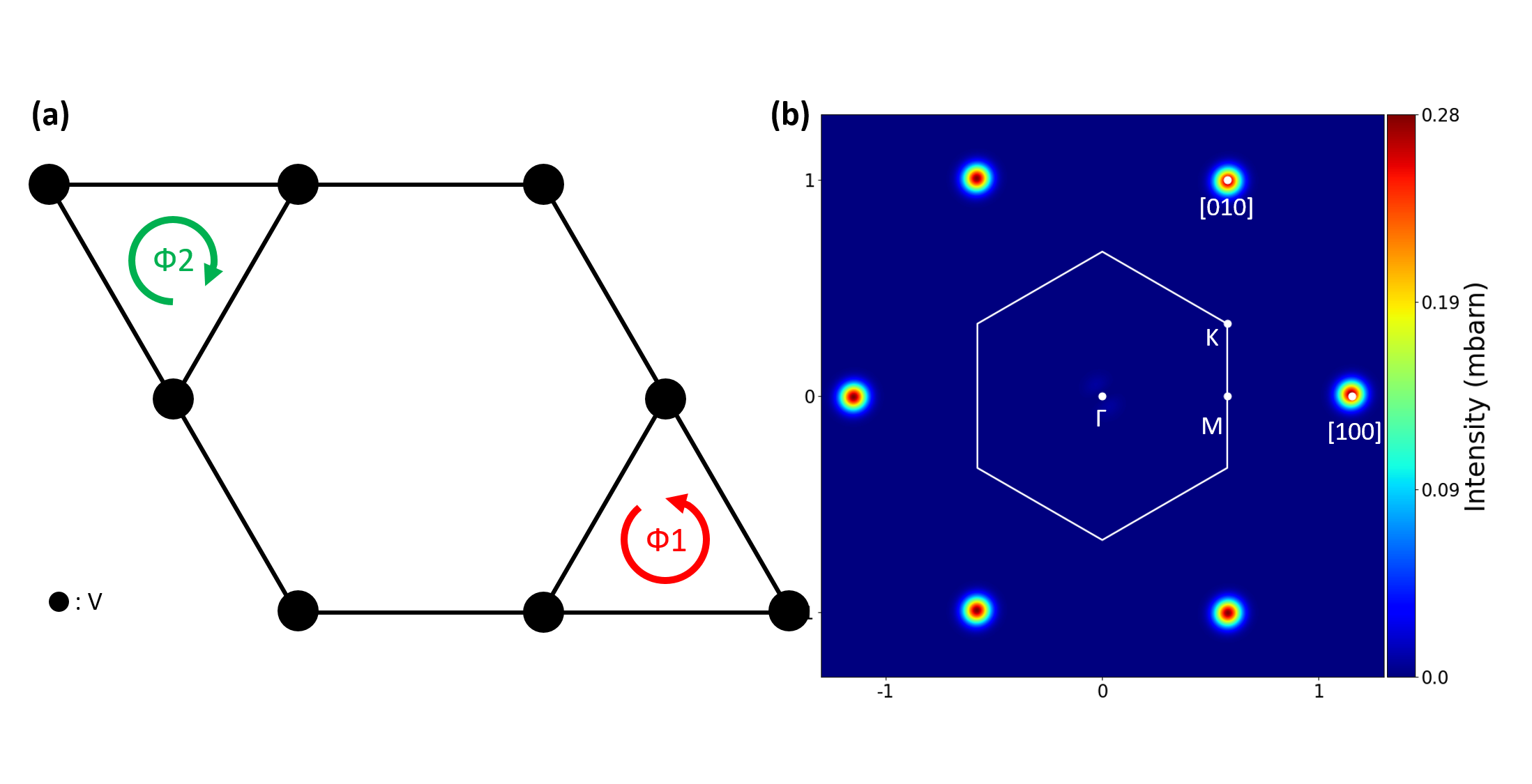} 
 \includegraphics[width=7cm]{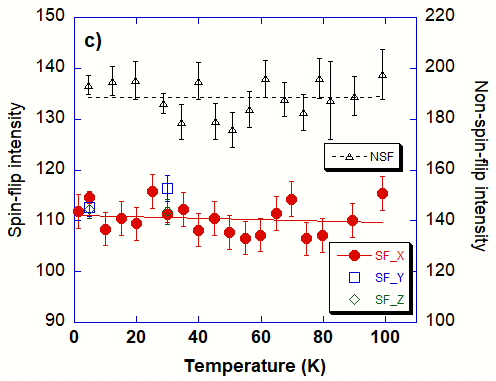} 
 \par\end{centering}
 \clearpage
\caption{\label{Fig9} \textbf{(a)} LCs configuration respecting the lattice symmetry  \textbf{(b)} Related calculated intensity with maxiumum at the Bragg position 
${\bf Q}=(1,0,0)$ \textbf{(c)} Temperature dependence of The Bragg peak ${\bf Q}=(1,0,0)$ of both spin-flip (with the three XYZ polarizations) (left scale) and non-spin-flip (right scale) intensities.  The intensities are all reported for a monitor M=1e6  corresponding  to a counting time of 317 sec although the counting time could reach 30 minutes/points for the most counted points.
 }
\end{figure}

First, no sizeable magnetic signal is observed at ${\bf M_1}$ compared to ${\bf M_2}$. If we are likely able to determine a weak magnetic scattering at ${\bf M_2}$ of $\sim$ 0.25 $\pm$ 0.1 mbarns, a similar magnetic amplitude at ${\bf M_1}$=(1/2,0,0) can be dismissed as the experimental limits are similar. Clearly, the magnetic signal is absent at the ${\bf M_1}$-points in the first Brillouin zone  questioning the theoretical LCs models in the kagome metals \cite{Feng21,Lin21,Zhou22} as none of the patterns  of Figs. \ref{Fig2}-\ref{Fig5} are consistent with the neutron data as a stronger intensity is always expected at ${\bf M_1}$ compared to ${\bf M_2}$. This suggests that the orbital  moments, if they exist, do not occur on vanadium hexagons.  This is for example the case in simpler models of moments on hexagons shown in Figs. \ref{Fig2}-\ref{Fig3} where the magnetic intensity at ${\bf M_1}$ and  ${\bf M_2}$ should be similar.  At least, it gives an upper limit for the putative magnetic moment. That value depends on the considered model, mainly depending on how many orbital moments are present in the 2x2 unit cell. Using that estimate and with the form factor given in the table \ref{TabfQ} which is $|f(Q)|^2$ =0.95 at 
${\bf M_1}= (1/2,0,0)$, applying Eq. \ref{FM-Feng21}, one gets for $I_m< 0.1$ mbarns at 5K $m_{hex}< 0.02 \mu_B$ for the model of Feng {\it et al}\cite{Feng21} (Fig. \ref{Fig4}). For the case of \cite{Lin21,Zhou22}  (Fig. \ref{Fig5}), 12 orbital moments contribute even if only three of them are actually clearly larger. In the model of Lin and Nandkishore \cite{Lin21}, one would get $m_{hex}< 0.012 \mu_B$ as there are more magnetic fluxes contributing to the structure factor. For the magnetic flux distribution proposed by Zhou and Wang \cite{Zhou22}, a similar orbital moment upper limit is obtained for the strongest orbital moment associated with the $\Phi_1$ flux.  In the simpler model of Fig. \ref{Fig2}, one needs further to consider the three magnetic domains: that leads to $m_{hex} < 0.016 \mu_B$ per formula unit in that case.

Among the different models discussed so far, the model of Feng {\it et al} \cite{Feng21} (Fig. \ref{Fig4}) is giving the higher upper limit of $m_{hex} < 0.02 \mu_B$ per formula unit which is well below  the experimental limit of STM with spin-polarized tips \cite{Li22} of about $\le 0.2 \mu_B$. Making the assumption that the magnetic moments are only carried out by the vanadium atoms, it gives a  tiny orbital moment of less than $0.01 \mu_B$  per vanadium atom. This value is substantially lower than what the different theories have been expecting\cite{Feng21,Lin21,Zhou22}. In particular, Zhou and Wang \cite{Zhou22} estimated an intrinsic thermodynamic orbital magnetization of 0.022 $\mu_B$ per vanadium atom whereas the experimental limit is $\sim 0.005 \mu_B$ per vanadium for this model. However, to determine the amplitude of the magnetic moment through the article, we assume a long range magnetic ordering. It is worth stressing that in case of finite magnetic correlation lengths, the extracted magnetic moment could be larger.  The low value of the measured orbital  moment might be as well related to the presence of impurities. It was indeed recently theoretically argued \cite{Nakazawa24} that the LCs state might be extremely sensitive to a small number of impurities and that the amplitude reduction originates from the non-local contribution from the itinerant circulation of electrons. 

Second, another conclusion is that  no magnetic signal occurs at various non zero $L$ values in both scattering planes. The only $L$ location where a signal can be extracted is at $L=0$ at the position  ${\bf M_2} = (1/2,1/2,0)$.  In particular, nothing is observed at $L$=1/2 that suggests no doubling of the unit cell along the $\bf c$-axis contrary to the discussion made from the $\mu$SR data \cite{Yu21}. We also do not see any magnetic signal at $L$=1/4 as it has been suggested for the CDW\cite{Ortiz21}. If  measurable, the magnetic signal is correlated in-phase from one hexagonal plane to the next one along the $\bf c$-axis. The magnetic ordering, if it exists, exhibits therefore in-phase three-dimensional correlations. 

Third,  we nevertheless detect a hint of orbital magnetism at the ${\bf M_2}$-points in the second Brillouin zones, suggesting a TRS breaking at low temperature in agreement with $\mu$SR experiments\cite{Mielke22,Yu21,Khasanov22,Guguchia23}. However, this should not be confused with the controversial reports of TRS breaking by Kerr effect \cite{Xu22,Saykin23} which is characteristic of a global TRS breaking whereas  neutron diffraction data  as well as muons are indicative of a local TRS breaking. Our finding of the strongest  magnetic intensity at the $\bf M_2$ points in the second Brillouin zone might be related to the recent ARPES observation  of pockets  \cite{Li23} (confirmed by quasi-particle interference in STM)  in the CDW state not observed in the first zone, but that appears at some (not all) of the predicted locations in the second zone in agreement with the theoretical prediction of the location of the pockets with a largest spectral weight \cite{Li23}.  

\begin{figure}
 \begin{centering}
 \includegraphics[width=14cm]{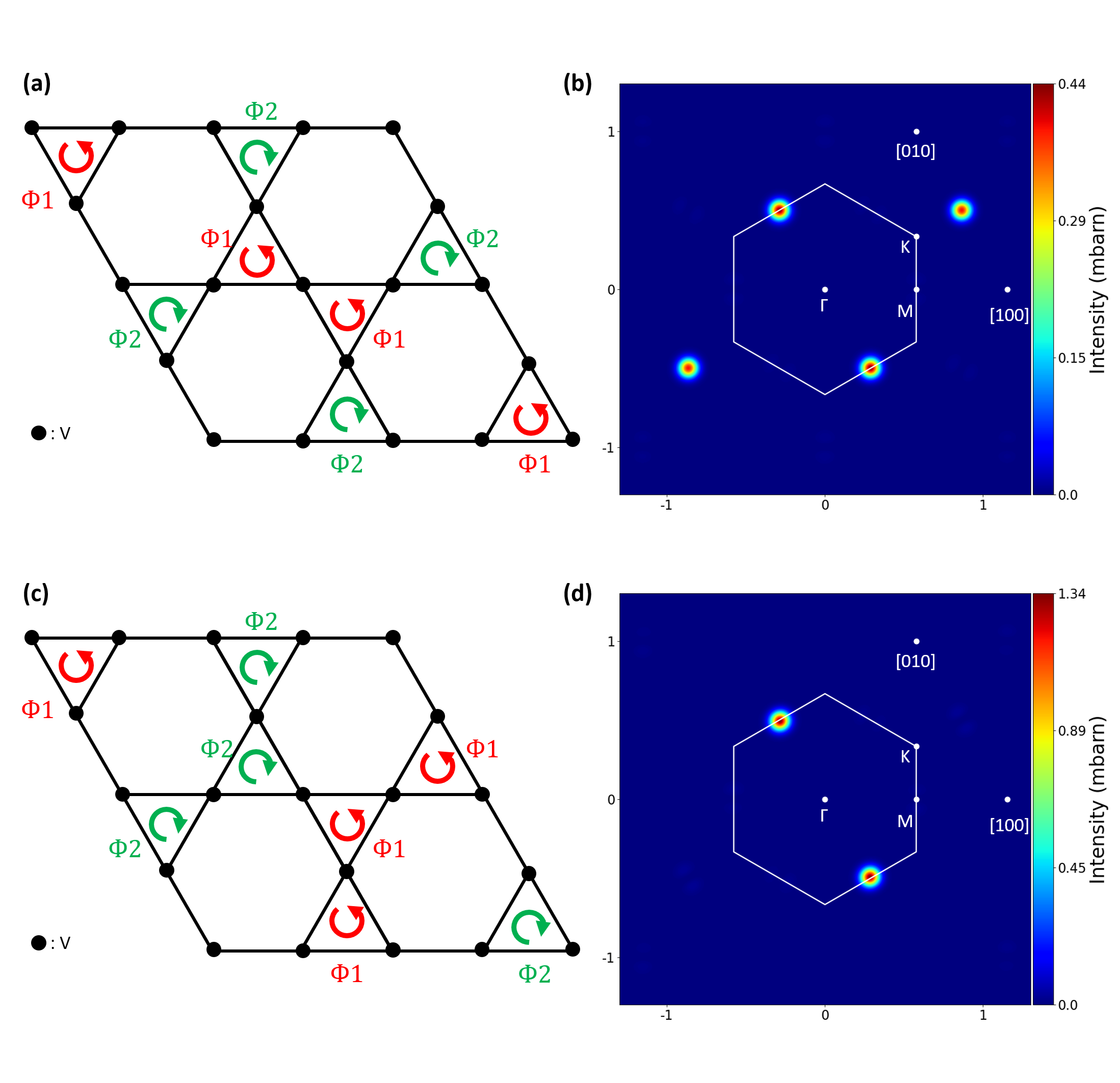}
 \par\end{centering}
 \clearpage
\caption{\label{Fig10}  \textbf{(a)} and \textbf{(c)}  Two LCs patterns with {\it only} moments on triangles in the 2x2 unit cell for an antiferromagnetic model that breaks $C_6$ rotation symmetry with $\Phi_1+\Phi_2$=0. \textbf{(b)} and \textbf{(d)}   Calculated magnetic intensity map in reciprocal space for both models. In  \textbf{(b)}, a larger intensity is found at  ${\bf M_2}$ compared to  ${\bf M_1}$. To  calculate the magnetic intensity in absolute units, the magnetic moment associated with $\Phi_1$ is set to 0.02 $\mu_B$. 
}
\end{figure}

Actually, such a result of an orbital magnetism at  ${\bf M_2} = (1/2,1/2,0)$ can be explained by LCs patterns flowing {\it only} on the triangles, as for instance it has been discussed  in the context of  $\mu$SR experiments \cite{Mielke22}. To be more explicit, one first considers the simplest LCs phase breaking $C_6$ discussed above for the hexagons (Fig. \ref{Fig2}). Instead of putting moment at the center of the hexagons, one now decorates all vanadium triangles of the 2x2 unit cell by magnetic moments with the symmetry of Fig. \ref{Fig2} and still with the constraint that the moment sum is zero ($4 \Phi_1+4\Phi_2=0$). Two situations, shown on Fig. \ref{Fig10}.a and \ref{Fig10}.c, are possible. One can repeat the calculation of the magnetic structure factor:  for the configuation of Fig. \ref{Fig10}.a, Eq  \ref{FM2x2} becomes,

\begin{equation}
 {\bf B( Q)}= 8 {m}_{tr}  \sin \pi H \sin \pi K \sin(\frac{\pi}{3}(2H+K))
\label{FM2x2tr}
\end{equation}

For the configuration of Fig. \ref{Fig10}.c, the factor $\sin(\frac{\pi}{3}(2H+K))$ in Eq. \ref{FM2x2tr} is changed by $\cos(\frac{\pi}{3}(2H+K))$. The calculated neutron intensity is reported in Fig. \ref{Fig10}.b and Fig. \ref{Fig10}.d for both configurations, respectively. One sees in Fig. \ref{Fig10}.d that the pattern of Fig. \ref{Fig10}.c is again inconsistent with our experimental data whereas Fig. \ref{Fig10}.b gives a noticeably larger intensity at  ${\bf M_2} = (1/2,1/2,0)$  compared to  ${\bf M_1} = (1/2,-1/2,0)$ that is favored by the pattern of Fig. \ref{Fig10}.a due to an additional decoration between triangles. This is clearly compatible with the experiment where only the magnetic intensity at  ${\bf M_2}$ is sizeable. 

\begin{figure}
 \begin{centering}
 \includegraphics[width=14cm]{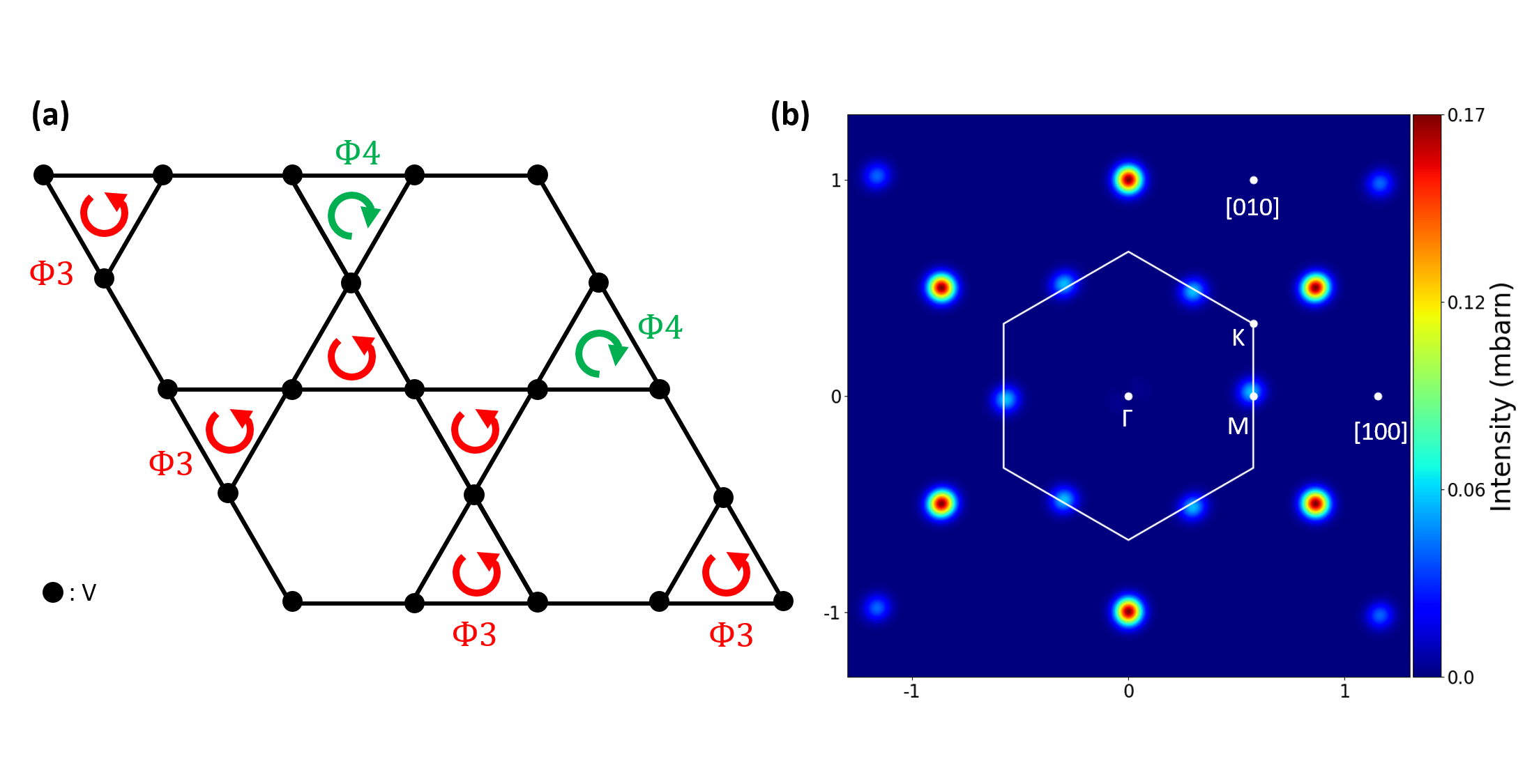}
 \par\end{centering}
 \clearpage
\caption{\label{Fig11}   \textbf{(a)} LCs configuration with moments  {\it only} on triangles in the 2x2 unit cell with a symmetry similar as the  proposed models \cite{Lin21,Zhou22} for the LCs in kagome metals.  It actually corresponds to the model of Zhou and Wang \cite{Zhou22} shown in Fig. \ref{Fig5} but with $\Phi_1=0$, $\Phi_2=0$ and $6 \Phi_3+2\Phi_4$=0. \textbf{(b)}   Calculated magnetic intensity map in reciprocal space with larger intensity at  ${\bf M_2}$  compared to ${\bf M_1}$. To  calculate the magnetic intensity in absolute units, the magnetic moment associated with $\Phi_3$ is set to 0.02 $\mu_B$.  }
\end{figure}

Using the above estimate of 0.25 mbarns  and with the form factor given in the table \ref{TabfQ} which is $|f(Q)|^2$ =0.85 at ${\bf M_2}$, one gets $m_{tr}=0.014 \pm 0.009 \mu_B$ per triangle at 5K for the model of Eq. \ref{FM2x2tr} where one has further to consider that we probe at ${\bf M_2} = (1/2,1/2,0)$ only 1/3 of the magnetic intensity due to the existence of magnetic domains. Actually, another LCs configuration, based on the proposed models \cite{Lin21,Zhou22} of Fig. \ref{Fig5}, can also account for the neutron data where {\it only} moments on triangles are considered with the constraint $6 \Phi_3+ 2\Phi_4=0$, that is akin to the structure shown in Fig. \ref{Fig11}.a. The related calculated neutron intensity of \ref{Fig11}.b suggests a map also compatible with the neutron data from which one can estimate at 5K $m_{tr}=0.024 \pm 0.013 \mu_B$ per triangle associated with the magnetic flux $\Phi_4$. That value is calculated from the related magnetic structure factor at ${\bf M_2}$, ${\bf B(\bf M_2) }= \frac{8}{3} {\bf {m}_{tr}}$. For the models of Fig. \ref{Fig10}.a and Fig. \ref{Fig11}.a, the calculated intensity at ${\bf M_1}$ is $\sim$ 3.6 weaker than at ${\bf M_2}$ which is then compatible with  the measured  upper limit at ${\bf M_1}$. This small magnetic moment,  deduced from the neutron data, sounds consistent with the $\mu$SR data \cite{Mielke22,Yu21,Khasanov22,Guguchia23} about the existence of a weak orbital moment ordering in kagome metals. For instance, only a field of 5 Gauss is extracted in ${\rm CsV_3Sb_5}$\cite{Mielke22} although we have not made any precise estimate as to how to convert the field seen at muon stopping site from our small reported magnetic moment. 

In conclusion, we searched for time-reversal symmetry breaking in the kagome superconductor  ${\rm CsV_3Sb_5}$, using polarized neutron diffraction.   At low temperature and particularly below the CDW temperature, we found within our experimental limit no magnetic scattering at ${\bf M_1}=(1/2,0,0)$-point in the first Brillouin zone although we have an indication of a weak signal at ${\bf M_2}=(1/2,1/2,0)$-point. This puts a very low limit on the expected magnetic orbital moment of less than 0.01 $\mu_B$ for the most popular models \cite{Feng21,Lin21,Zhou22} proposed for the chiral flux state breaking time-reversal symmetry that are discussed in Fig. \ref{Fig4}-\ref{Fig5}. These results are compatible with loop currents flowing {\it only} on the vanadium triangles. From these data,  we could determine a very weak orbital magnetic moment of  the order of $m_{tr} \sim 0.02 \pm 0.01 \mu_B$ per vanadium triangle. Improving the experiment in the future would require a sample with a much better mosaicity and a larger mass. The kagome metals are a nice opportunity to show that loop currents can exist over a wider range of quantum materials.  Many other materials should belong to that category, particularly low dimensional and unconventional superconductors. 

% ========================================Acknowledgements
 \noindent \textbf{Acknowledgments}

 We thank Chandra Varma and Ziqiang Wang for stimulating discussions on loop currents in kagome metals. We acknowledge private communications from Yu Ping Lin and Rahul Nankishore who provide the detailed flux patterns corresponding to their model \cite{Lin21}. We thank Victor Bal\'edent, Quentin Faure  and Paul McClarty for critical reading of the manuscript. We are grateful to the full IRIG/D-phy/MEM/MDN group of CEA Grenoble who helped us to mount the different experimental setups. We also thank Wolfgang  Schmidt for assistance during one experiment. We also acknowledge financial support from the F\'ed\'eration Fran\c caise de Diffusion Neutronique (2FDN). The neutron scattering and basic materials synthesis work at Rice University are supported by the US DOE, Basic Energy Sciences (BES), undergrant no. DE-SC0012311 and by the Robert A. Welch Foundation under grant no. C-1839, respectively.

% Corresponding author: Philippe Bourges.

 %\bibliographystyle{nature}
\bibliography{references}

\newpage

% ========================================Fig 1

% ========================================Fig 2

% ========================================Fig 3

% ========================================Fig 4

% ========================================Fig 5

% ========================================Fig 6

% ========================================Fig 7

% ========================================Fig 8

% ========================================Fig 9

% ========================================Fig 10

% ========================================Fig 11

\end{document}